\documentclass[12pt]{iopart}

\usepackage{cite}
\usepackage{amssymb,amsfonts}
\usepackage{algorithmic}
\usepackage{graphicx}
\expandafter\let\csname equation*\endcsname\relax
\expandafter\let\csname endequation*\endcsname\relax
\usepackage{amsmath}
\usepackage{bm}
\usepackage{textcomp}
\usepackage{multirow}
\usepackage{lscape}
\usepackage{booktabs}
\usepackage{bbm}
\usepackage[utf8]{inputenc}
\usepackage{times}
\usepackage{soul}
\usepackage{url}
\usepackage{multicol}
\usepackage[ruled,vlined]{algorithm2e}
\usepackage{lscape}
\usepackage{rotating}

\usepackage{wrapfig}
\usepackage{colortbl}

\definecolor{gray}{rgb}{0.5,0.5,0.5}
\definecolor{darkergreen}{RGB}{21, 152, 56}
\newcommand{\tabincell}[2]{\begin{tabular}{@{}#1@{}}#2\end{tabular}}

\begin{document}

\title[]{Transfer learning on electromyography (EMG) tasks: approaches and beyond}
% \author{P J Smith, T M Collins, 
% R J Jones$^3$\footnote{Present address:
% Department of Physics, University of Bristol, Tyndalls Park Road, 
% Bristol BS8 1TS, UK.} and Janet Williams$^3$}

\author{Di Wu, Jie Yang, and Mohamad Sawan}

\address{Center of Excellence in Biomedical Research on Advanced Integrated-on-chips Neurotechnologies, School of Engineering, Westlake University, Hangzhou, 310024, China}

\address{Institute of Advanced Technology, Westlake Institute for Advanced Study, Hangzhou, 310024, China}

\ead{\mailto{yangjie@westlake.edu.cn}, \mailto{Sawan@westlake.edu.cn}}

\vspace{10pt}
% \begin{indented}
% \item[]August 2017
% \end{indented}
\begin{abstract}
% Deep learning on graphs  on a variety of tasks, while such success relies heavily
% on the massive and carefully labeled data. However, precise annotations are generally very expensive and time-consuming. To address
% deep neural networks (DNN) 
% this problem, self-supervised learning (SSL) is emerging as a new paradigm for extracting informative knowledge through well-designed
% pretext tasks without relying on manual labels.
\it{Objective}. Machine learning on electromyography (EMG) has recently achieved remarkable success on a variety of tasks, while such success relies heavily on the assumption that the training and future data must be of the same data distribution. However, this assumption may not hold in many real-world applications. Model calibration is required via data re-collection and label annotation, which is generally very expensive and time-consuming. To address this problem, transfer learning (TL), which aims to improve target learners' performance by transferring the knowledge from related source domains, is emerging as a new paradigm to reduce the amount of calibration effort. \it{Approach}. In this survey, we assess the eligibility of more than 
fifty published peer-reviewed representative transfer learning approaches for EMG applications. \it{Main results}. Unlike previous surveys on purely transfer learning or EMG-based machine learning, this survey aims to provide an insight into the biological foundations of existing transfer learning methods on EMG-related analysis. In specific, we first introduce the physiological structure of the muscles and the EMG generating mechanism, and the recording of EMG to provide biological insights behind existing transfer learning approaches. Further, we categorize existing research endeavors into data based, model based, training scheme based, and adversarial based. \it{Significance}. This survey systematically summarizes and categorizes existing transfer learning approaches for EMG related machine learning applications. In addition, we discuss possible drawbacks of existing works and point out the future direction of better EMG transfer learning algorithms to enhance practicality for real-world applications.

\end{abstract}

\vspace{2pc}

\section{Introduction}
\label{sec:introduction}
The human motor control system is a complex neural system that is crucial for daily human activities. One way to study the human motor control system is to record the signal due to muscle fiber contractions associated with human motor activities by means of either inserting needle electrodes into the muscles or attaching electrodes onto the surface of the skin. The signal obtained is referred to as electromyography (EMG). Given the location of the electrodes, EMG is further divided into surface EMG (sEMG) and
intramuscular EMG (iEMG). Advancement in the analysis of EMG and machine learning has recently achieved remarkable success enabling a wide variety of applications, including but not limited to rehabilitation with prostheses~\cite{ghassemi2019development}, hand gesture recognition~\cite{qi2020surface} and human-machine interfaces (HMIs)~\cite{saponas2009enabling}.

The current success of applying deep learning onto EMG related tasks is largely confined to the following two assumptions, which are usually infeasible when it comes to real-world EMG related scenarios: 

\begin{itemize}
\item[1)] \textbf{Sufficient amount of annotated training data.} The growing capability and capacity of deep neural networks (DNN) architectures are associated with million-scale labeled data~\cite{krizhevsky2012imagenet, vaswani2017attention}. Such high quality abundant labeled data are often limited, expensive, and inaccessible in the domain of EMG analysis. On the one hand, EMG data annotation requires expert knowledge. On the other hand, EMG data acquisition process is a highly physical and time-consuming task that requires several days of collaboration from multiple parties~\cite{pizzolato2017comparison}.
\item[2)] \textbf{Training data and testing data are independent and identically distributed (i.i.d).} The performance of the model is largely affected by the distribution gap between the training and testing datasets. The testing data might also refer to the data generated during actual application usage after model deployment. Take hand gesture recognition, for example. The model is only capable of giving accurate predictions with the exact same positioning of the forearm of the test subject and the exact placement of the electrodes.
\end{itemize}

As the distribution of data changes, models based on statistics need to be reconstructed with newly collected training data. In many real-world applications, it is expensive and impractical to recollect a large amount of training data and rebuild the models each time a distribution change is observed. Transfer learning (TL), which emphasizes the transfer of knowledge across domains, emerges as a promising machine learning solution for solving the above problems. The notion of transfer learning is not new, Thorndike \textit{et al.}~\cite{woodworth1901influence} suggested that the improvement over one task is beneficial to the efficiency of learning other tasks given the similarity exists between these two tasks. In practice, a person knowing how to ride a bicycle can learn to ride a motorcycle faster than others since both tasks require balance keeping. However, transfer learning for EMG related tasks has only been gaining attention with the recent development of both DNN and HMIs. Existing surveys provide an overview of DNN for EMG-based human machine interfaces~\cite{xiong2021deep}, and transfer learning in general for various machine learning tasks~\cite{zhuang2020comprehensive}. This survey focuses on the intersection of machine learning for EMG and transfer learning via EMG biological foundations, providing insights into a novel and growing area of research. Besides the analysis of recent deep learning works, we make an attempt to explain the relationships and differences between non-deep learning and the deep models, for these works usually share similar intuitions and observations. Some of the previous non-deep learning works contain more biological significance that can inspire further DNN-based research in this field. To consolidate these recent advances, we propose a new taxonomy for transfer learning on EMG tasks, and also provide a collection of predominant benchmark datasets following our taxonomy.

% \begin{figure}[t]
% \centerline{\includegraphics[width=\columnwidth]{pics/motivation4.pdf}}
% \caption{Comparison among patient-independent, patient-specific, and our training scheme. Patient-independent training scheme (blue region) trains a model using data pooled from all subjects. Patient-specific training scheme (yellow region) trains a unique model for each subject. Our training scheme (red region) first trains a P network using data pooled from $k-1$ subjects in the first training stage. Then during the second training stage, we randomly initialize a C network with the same structure as the P network. By applying knowledge distillation to the P network and C network using the $k^{th}$ subject's data, we generate a distilled C network that enables better prediction performance.}
% \label{fig_intro}
% \end{figure}

The main contributions of this paper are :

\begin{itemize}
\item Over fifty representative up-to-date transfer learning approaches on EMG analysis 
are summarized with organized categorization, presenting a comprehensive overview to the readers.
\item Delve deep into the generating mechanisms of EMG and bridge transfer learning practices with the underlying biological foundation.
\item Point out the technical limitations of current research and discuss promising directions on transfer learning on EMG analysis to propose further studies.
\end{itemize}

The remainder of this paper is organized as follows. We introduce in section \ref{sec:preliminaries} the basics of transfer learning, generation and acquisition of EMG and EMG transfer learning scenarios. In Section \ref{sec:tl_emg}, we first provide the categorization of EMG transfer learning based on existing works and then introduce in detail. We also give a summary of common used dataset in Section \ref{sec:dataset}. Lastly, we discuss existing methods and the future research direction of EMG transfer learning.
%  In section \ref{sec:method}, we provide a detailed description of our proposed training scheme. The performance of this proposed training scheme is evaluated in Section \ref{sec:results}. We then provide a discussion on results, limitations, and future work in Section \ref{sec:discussion}. Finally, the paper is concluded in Section \ref{sec:conclusion}.
\section{Preliminaries}
\label{sec:preliminaries}
This section introduces the definitions of transfer learning, related concepts, and also the basics of EMG, from how EMG signal is generated to how EMG signal is recorded. We also summarize possible transfer scenarios in this section. 

\subsection{Transfer Learning}
We first give the definitions of a “domain” and a “task”, respectively. Define $\mathcal{D}$ to be a domain which consists of a feature space $\mathcal{X}$ and a marginal probability distribution $P(X)$, where $X$ is a set of data samples $X=[x_i]_{i=1}^{n}$. In particular, if two domains have different feature spaces or marginal probability distributions, they differ from each other. Given a domain $\mathcal{D} = \{ \mathcal{X},\ P(X)\}$, a task is then represented by $\mathcal{T} = \{ \mathcal{Y}, f(\cdot)\}$ where $f(\cdot)$ denotes the objective prediction function and $ \mathcal{Y}$ is the label space associated with $\mathcal{X}$. From the probability point of view, $f(x)$ can also be regarded as conditional probability distribution $P(y|x)$. Two tasks are considered different if they have different label spaces of different conditional probability distributions. Then, transfer learning can be formally defined as follows:

\textit{Definition 1 (Transfer Learning)}: Given a source learning task $\mathcal{T}_S$ based on a source domain $\mathcal{D}_S$, transfer learning aims to help improve the learning of the target objective prediction function $f_{\mathcal{T}}(x)$ of the target task $\mathcal{T}_S$ based on the target domain $\mathcal{D}_T$, given that $\mathcal{D}_T \neq \mathcal{D}_S$ or $\mathcal{T}_S \neq \mathcal{T}_T$.

The above definition could be extended to multiple domains and tasks for both source and target. In this survey, we only consider the case where there is one source domain $\mathcal{D}_S$, and one target domain $\mathcal{D}_T$, as by far this is the most intensively studied transfer setup of the research works in the literature. Based on different setups of the source and target domains and tasks, transfer learning could be roughly categorized into \textit{inductive transfer learning}, \textit{transductive transfer learning} and \textit{unsupervised transfer learning}~\cite{pan2009survey}.

\textit{Definition 2 (Inductive Transfer Learning)}: Given a transfer learning task $(\mathcal{D}_S, \mathcal{T}_S, \mathcal{D}_T, \mathcal{T}_T, f_{\mathcal{T}}(x)) $. It is a \textit{inductive transfer learning} task where the knowledge of $(\mathcal{D}_S$ and $\mathcal{T}_S$ is used to improve the learning of the target objective prediction function $f_{\mathcal{T}}(x)$ when $\mathcal{T}_S \neq \mathcal{T}_T$.

The target objective predictive function can be induced by using a few labeled data in the target domain as the training data.

\textit{Definition 3 (Transductive Transfer Learning)}: Given a transfer learning task $(\mathcal{D}_S, \mathcal{T}_S, \mathcal{D}_T, \mathcal{T}_T, f_{\mathcal{T}}(x)) $. It is a \textit{transductive transfer learning} task where the knowledge of $\mathcal{D}_S$ and $\mathcal{T}_S$ is used to improve the learning of the target objective prediction function $f_{\mathcal{T}}(x)$ when $\mathcal{D}_S \neq \mathcal{D}_T$ and $\mathcal{T}_S = \mathcal{T}_T$.

For \textit{transductive transfer learning}, the source and target tasks are the same, while the source and target domain vary. Similar to the setting of transductive learning of traditional machine learning\cite{joachims1999transductive}, \textit{transductive transfer learning} aims to make the best use of the given unlabeled data in the target domain to adapt the objective predictive function learned in the source domain, minimizing the expected error on the target domain. It is worth to notice that \textit{domain adaptation} is a special case where $\mathcal{X}_{S} = \mathcal{X}_{T}$, $\mathcal{Y}_{S} = \mathcal{Y}_{T}$, $P_{S}(y|X) \neq P_{T}(y|X)$ and/or $P_{S}(X) \neq P_{T}(X)$.

\textit{Definition 4 (Unsupervised Transfer Learning)}: Given a transfer learning task $(\mathcal{D}_S, \mathcal{T}_S, \mathcal{D}_T, \mathcal{T}_T, f_{\mathcal{T}}(x)) $. It is an \textit{unsupervised transfer learning} task where the knowledge of $\mathcal{D}_S$ and $\mathcal{T}_S$ is used to improve the learning of the target objective prediction function $f_{\mathcal{T}}(x)$ with $\mathcal{Y}_{S}$ and $\mathcal{Y}_{T}$ not observed.

Based on the above definition, no data annotation is accessible in both the source and target domain during training. There has been little research conducted on this setting to date, given its fully unsupervised nature in both domains.  

\begin{figure*}[t]
\centerline{\includegraphics[width=\columnwidth]{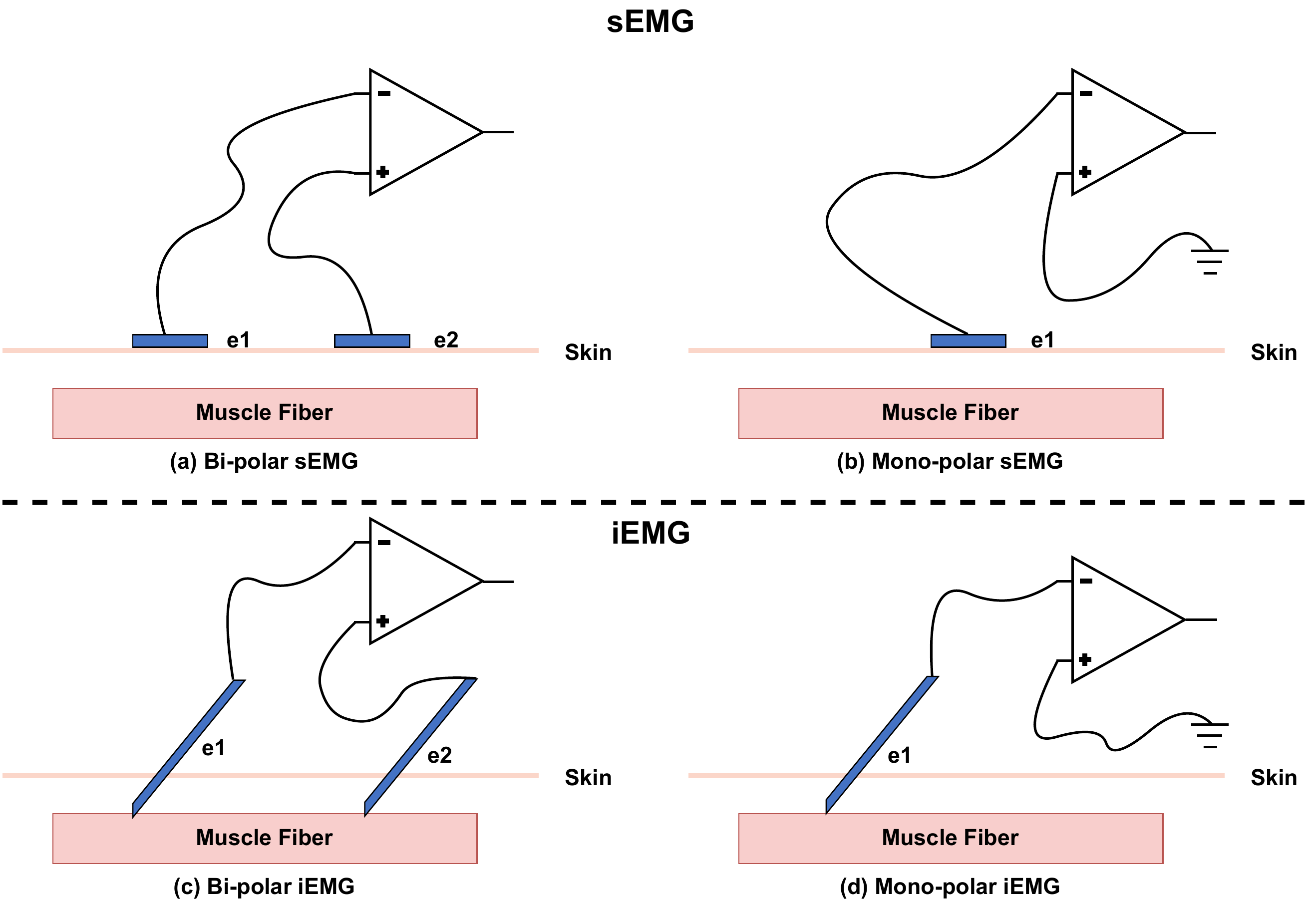}}
	\caption{
	Demonstration of EMG acquisition. The sEMG acquisition configuration is shown above the dotted line, with iEMG acquisition configuration shown below the dotted line. The triangle represents an amplifier. For the bi-polar setup as in (a) and (c), two electrodes are placed on the skin surface or inserted into muscle fibers penetrating the skin surface. (b) and (d) show the case of a mono-polar setup with one electrode attached to the skin or muscle fiber and the other electrode connected to the ground or a reference point with no EMG (bones).}
	\label{fig:acq}
\end{figure*}
\subsection{EMG Basics}
\label{sec:emg_basics}
\paragraph{Motor Unit Action Potential.} A \textit{motor unit} (MU) is defined as one motor neuron and the muscle fibers that it innervates. During the contraction of a normal muscle, the muscle fibers of a motor unit are activated by its associated motor neuron. The membrane depolarization of the muscle fiber is accompanied by ions movement and thus generates an electromagnetic field in the vicinity of the muscle fiber. The detected potential or voltage within the electromagnetic field is referred to as the fiber action potential. The amplitude of the fiber action potential is related to the diameter of the corresponding muscle fiber and the distance to the recording electrode. It is worth noticing that MU, by definition, refers to the anatomical motor unit where the \textit{functional motor unit} is of more research interest when it comes to real-world applications. The functional motor unit can be defined as a group of muscle fibers whose action potentials occur within a very short time (two milliseconds). Intuitively, one could consider a functional motor unit as a group of muscle fibers that contract for one unified functionality. From this point on, MU refers to a functional motor unit unless otherwise specified. A Motor Unit Action Potential (MUAP) is defined as the waveform consisting of the superimposed (both temporally and spatially) action potentials from each individual muscle fiber of the motor unit. The amplitude and shape of the MUAP is a unique indicator of the properties of the MU (functionality, fiber arrangement, fiber diameter, etc.). MUs are repeatedly activated so that muscle contraction is sustained for stable motor movement. The repeated activation of MUs generates a sequence of MUAPs forming a Motor Unit Action Potential Train (MUAPT).

\paragraph{Signal Recording.}  

Based on the number of electrodes used during the recording of MUAPT, the recording techniques could be divided into mono-polar and bi-polar configurations. As shown in Figure \ref{fig:acq}, based on whether the electrodes are inserted directly into the muscles or placed on the surface of the skin, the collected signal is referred to as intramuscular EMG (iEMG) or surface EMG (sEMG), respectively. If muscle fibers belonging to multiple MUs are within the vicinity of the electrode, all MUAPTs from different MUs will be detected by the electrode. A thin and sharp needle shaped electrode is quickly and smoothly inserted into the targeted muscle during iEMG acquisition~\cite{daube2009needle}. iEMG is considered to have good spatial resolution due to the small diameter (around 0.5 mm) of the needle electrode. Individual MUAPTs could be identified by visualization. However, the effectiveness of the process of iEMG acquisition is highly dependent on the skill of the electrodiagnostic physician. Moreover, the punctuation procedure bears the risks such as skin infection, severe bleeding, and muscle irritation. sEMG, on the other hand, is a non-invasive analysis tool for the human motor system places electrodes on the surface of the skin~\cite{hermens2000development}. Given the different diameters of the electrode, sEMG is composed of MUAPTs from MUs from the same layer or deep layers, leading to a poor spatial resolution as compared to iEMG. sEMG is widely adopted for Human-Computer Interface (HCI) due to the major advantage of its ease of use and non-invasive nature.
\begin{figure*}[t]
\centerline{\includegraphics[width=\columnwidth]{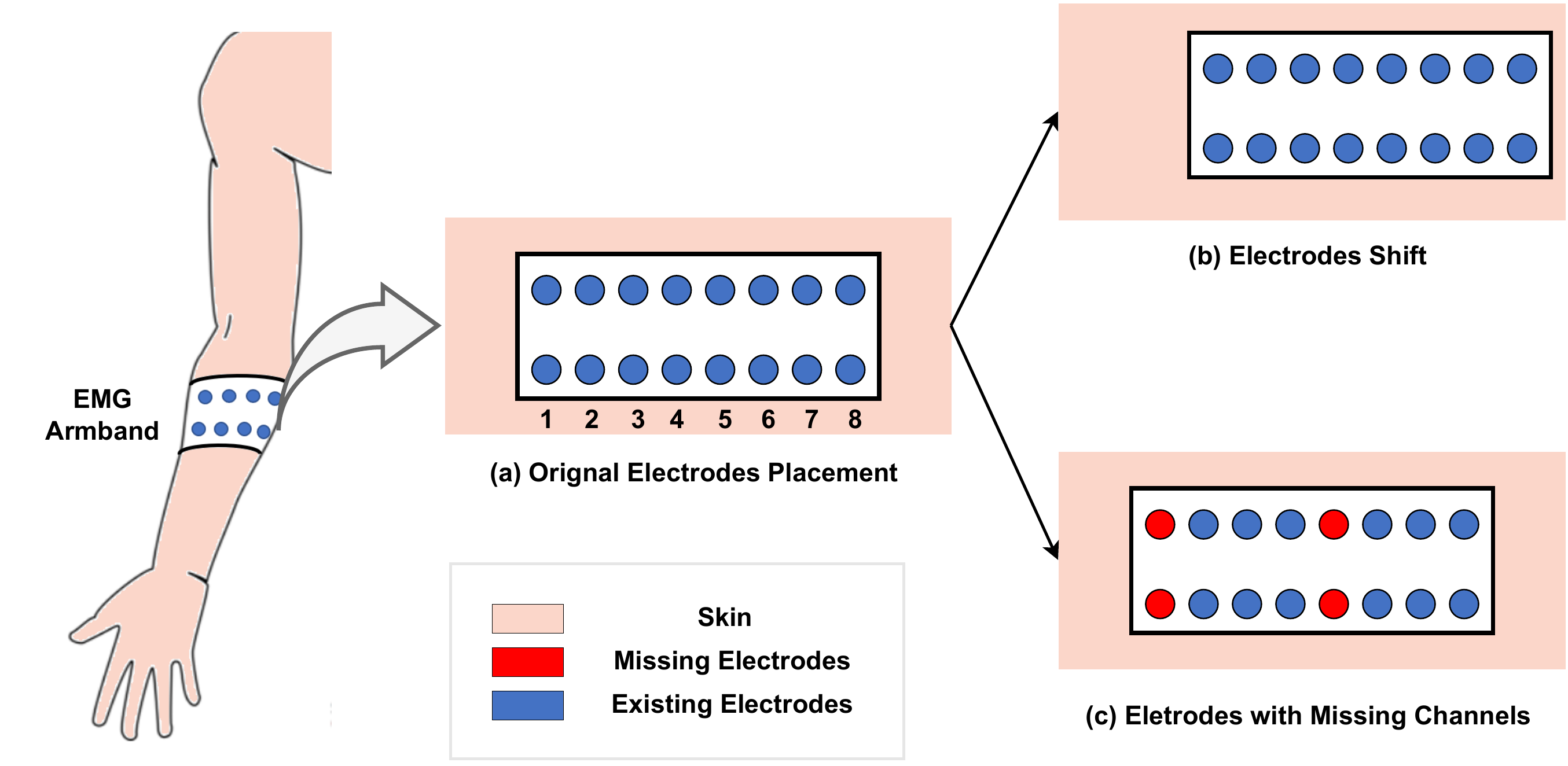}}
	\caption{
	Illustration of electrode variation. The left-hand side shows an EMG acquisition armband put on the forearm of a subject. (a), (b) and (c) are the net of the armband and the corresponding skin underneath. Colored circles represent electrodes, with two vertically placed electrodes being one bi-polar channel. (a) demonstrates the original placement of an eight-channel bi-polar EMG collecting armband on the surface of the skin. (b) shows a shifted placement of the electrodes on the skin compared to (a). (c) is the case where electrode placement is the same as (a), but some channels are missing due to any reason.}
	\label{fig:shift}
\end{figure*}
\subsection{Transfer Scenarios of EMG}
\label{sec:trans_setting}
Based on various factors in real usage scenarios that cause a difference between the source domain and the target domain, we summarize transfer settings in EMG based applications as follows:
\begin{itemize}
\item[1)] \textbf{Electrodes Variation.} Electrode variation could be categorized into electrodes placement shift and channel variation. Channel variation refers to the situation where some channels are missing during actual use as compared to the number of channels while recording EMG for model training. The placement of electrodes plays a crucial role in EMG applications. However, electrode shift is inevitable from wearing and taking off EMG acquisition devices whether in the form of armband~\cite{xiong2021deep} or sockets\cite{fleming2021myoelectric}. Figure \ref{fig:shift} provides a visualization of electrode variation in the case of an eight-channel EMG armband acquisition device. Consider the task of hand gesture and source domain associated with data collected with electrode placement shown in Figure \ref{fig:shift}(a). A transfer learning setting is formed with the target domain consisting of the same task and data collected with electrode placement shown in Figure \ref{fig:shift}(b) or with missing channels as in Figure \ref{fig:shift}(c).
\item[2)] \textbf{Inter-subject.} EMG signals have substantial variation across individuals. The variation might come from a different distribution of subcutaneous fat, muscle fiber diameter, and way of performing force. Inter-subject transfer refers to the scenario where data collected from one subject or other subjects is utilized to calibrate the target objective function on a new subject. The task and acquisition devices are assumed to be the same across individuals.
\item[3)] \textbf{Inter-session.} In real-world applications, models are built with data collected from previous sessions and applied to new sessions. Data distribution varies across sessions due to reasons such as a different way of performing gestures, variation in electrode placement, or simply muscle fatigue. Inter-session transfer refers to the scenario where data collected from previous sessions is utilized to calibrate the target objective function in a new session. The task, acquisition device, and subject are assumed to be the same across sessions.
\item[4)] \textbf{Modality Variation.} Modality transfer refers to the scenario where data collected on one or a few modalities is utilized to calibrate the target objective function on another or other modalities. The task and subject are assumed to be the same, while devices vary due to modality variation. For the same or relevant tasks, it is possible to utilize the knowledge learned from one modality and facilitate the performance of the objective prediction function on another modality. For example, the transfer learning due to modality variation could be between neurophysiological signals (EEG and EMG)~\cite{wu2022neuro2vec}.
\item[5)] \textbf{Extensive Learning.} Extensive learning refers to the transfer scenario new input data (target domain) extends either the data or/and the task of the source domain. For instance, the task of the source domain is a $C$ class classification problem while data collected in the target domain is of $C+K$ classes where $K$ additional classes are incrementally added. The acquisition device and subject are assumed to be the same for both domains.
\end{itemize}
\begin{figure*}[t]
\centerline{\includegraphics[width=\columnwidth]{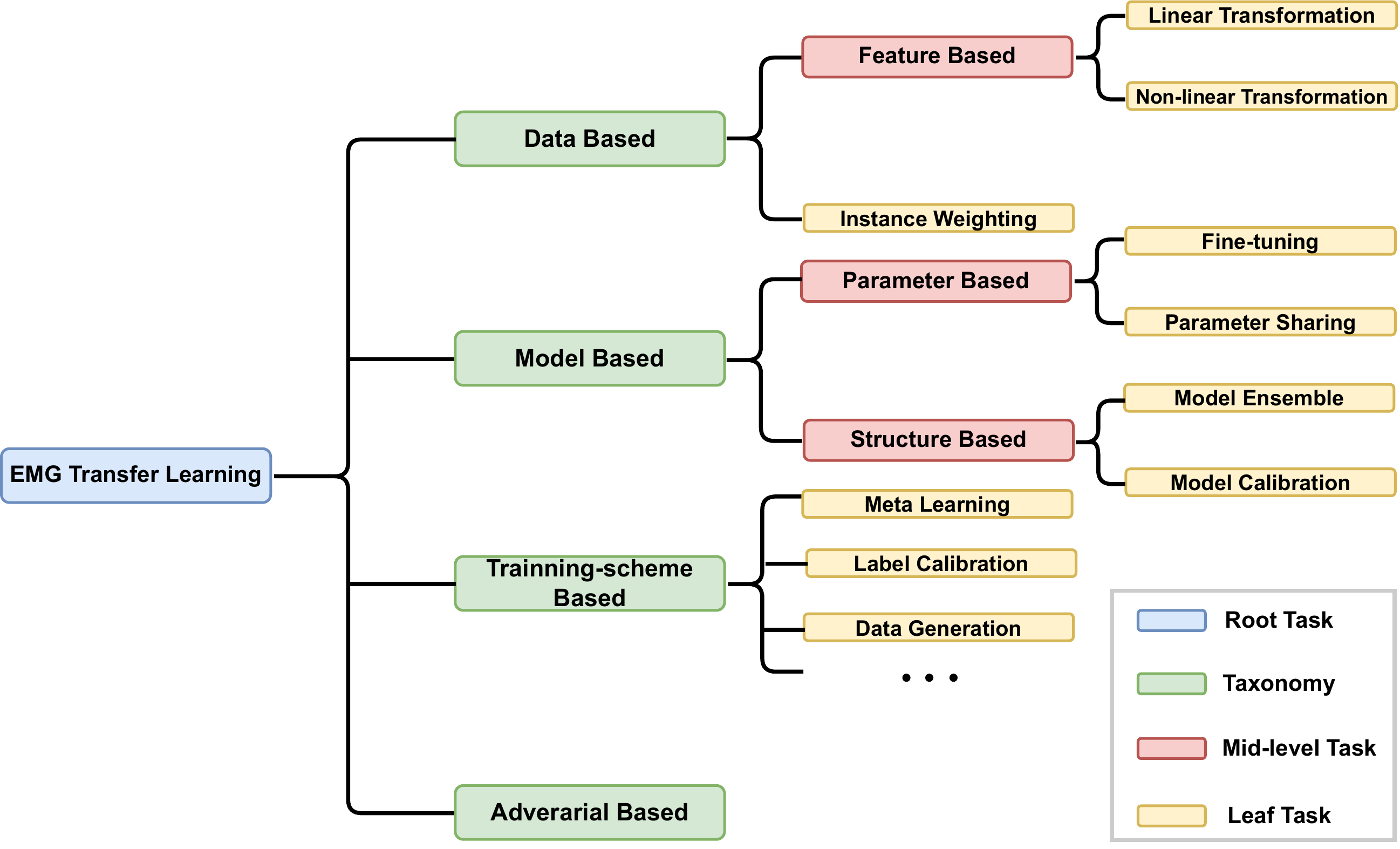}}
	\caption{
	Overview of categorization of transfer learning on EMG analysis. 
	}
	\label{fig:tax}
\end{figure*}
\section{Transfer Learning in EMG Analysis}
\label{sec:tl_emg}
In the previous section, we introduced basic concepts on transfer learning on general and EMG generating mechanisms along with recording techniques. These preliminaries shed insights on the underlying principles of recent progress in the area of transfer learning on EMG. In this section, we construct a categorization that best summarizes existing research endeavors of transfer learning in EMG analysis. As shown in Figure~\ref{fig:tax}, we categorize existing works in EMG related transfer learning into four lines, \textit{i.e.}, data-based approaches, model-based approaches, training scheme based approaches, and adversarial-based approaches. Considering whether the approach weights the data instance or apply feature transformation, we further divide data-based approaches into feature based methods and instance weighting approaches. In similar ways, we further divide model-based approaches into parameter-based and structure-based. Even further, we divide parameter-based methods into parameter sharing and fine-tuning while splitting structure based methods into the model ensemble and model calibration. Besides model-based and data-based interpretation, some transfer strategies are based on specially designed training schemes or adversarial training.

\subsection{Data-based Perspective}
Data-based transfer learning approaches aim to reduce the data distribution difference between the source domain and target domain via data transformation and adjustment. From a data perspective, two approaches are generally employed in order to accomplish the knowledge transfer objective, namely instance weighting and feature based transformation. According to the strategies illustrated in Figure \ref{fig:tax}, we present some most related approaches.

\subsubsection{Instance Weighting} Consider a special case of domain adaptation where $P_{S}(y|X) = P_{T}(y|X)$ and $P_{S}(X) \neq P_{T}(X)$ which is referred to as covariate shift~\cite{shimodaira2000improving}. Consider the transfer scenarios that we introduced in Section \ref{sec:trans_setting}, collecting abundant data in the target domain is often prohibitive, and thus target domain instances are limited. A natural solution is to assign weights to partial instances from the source domain so that these source domain instances can be used along with limited target domain data. Huang \textit{et al.} proposed Kernel Mean Matching (KMM)~\cite{huang2006correcting} to estimate the instance weights by matching the means of the target and source domain in a Reproducing Kernel Hilbert Space (RKHS). The weighted instances from the source domain are combined with labeled target domain instances to train the target objective prediction function. Li \textit{et al.}~\cite{li2021improvement} proposed to use TrAdaBoost~\cite{10.1145/1273496.1273521} along with Support Vector Machine (SVM) to improve the motion recognition performance under inter-session scenario. In specific, they first apply TrAdaBoost to weight EMG data of day one and train a target classifier with weighted EMG from day one and EMG collected from another day. TrAdaBoost iteratively adjusts the weights of instances to decrease the negative effect of the instances on the target learner. TrAdaBoost is largely inspired by a boosting algorithm called AdaBoost~\cite{freund1997decision}. AdaBoost iteratively trains weak classifiers with updated weights. The weighting mechanism of AdaBoost is the misclassified instances are given more attention during the training of the next weak learner in the following iteration. The weighting mechanism of TrAdaBoost is to reduce the distribution difference between the source domain and the target domain.
\subsubsection{Feature Based Strategy}
\label{sec:feature}
Feature-based approaches map each original feature into a new feature representation either by linearly transforming the original feature or non-linearly transforming the original feature to enable knowledge transfer. 
\paragraph{Linear Transformation.}
Lin \textit{et al.}~\cite{lin2020normalisation} proposed a normalization based approach called Referencing Normalisation to reduce the distribution difference among domains for inter-subject sEMG-based hand gesture classification. In specific, data from the source domain are mapped to the range of the target domain data:
\begin{equation}
\tilde{X_S} = \frac{X_S-min(X_S)}{max(X_S)-min(X_S)} * (max(X_T)-min(X_T)) + min(X_T),
\label{eq3}
\end{equation}
where $\tilde{X_S}$ is the transformed source domain data.

In addition to directly applying a linear transformation to normalize the data to the target domain range, authors~\cite{7302056,6985518,ameri2019deep, kanoga2018assessing} attempted to reduce the distribution gap based on statistical features such as covariance and mean. Conventional classifiers such as Linear Discriminant Analysis (LDA)~\cite{kaufmann2010fluctuating}, Quadratic Discriminant Analysis (QDA)~\cite{hastie2009elements} and Polynomial Classifier (PC)~\cite{al2006surface} are commonly adopted for sEMG classification tasks. The covariance matrix, mean vector, and the prior are the discriminant variables of LDA and QDA classifiers. Define $\mathbf{\Sigma}_S, \mathbf{\Sigma}_T, \mathbf{\mu}_S, \mathbf{\mu}_T$ to be the covariance matrices and mean vectors of data from the source domain and target domain, respectively. The transfer learning process of LDA and QDA based linear classifiers could be defined with a convex interpolation:
\begin{subequations}
\begin{align}
\tilde{\mathbf{\Sigma}} = (1-\alpha) * \mathbf{\Sigma}_S + \alpha * \mathbf{\Sigma}_T \\
\tilde{\mathbf{\mu}} = (1-\beta) * \mathbf{\mu} + \beta * \mathbf{\mu}_T,
\end{align}
\end{subequations}
where $\alpha, \beta \in [0,1]$ are the trade-off parameters to balance the knowledge from the source and target domain, $\tilde{\mathbf{\Sigma}}$ and $\tilde{\mu}$ represent the adapted covariance and mean vector. The optimal value for $\alpha$ and $\beta$ are set empirically or via grid search with a fixed step size. Liu \textit{et al.}~\cite{6985518} also proposed to use transfer learning on PC for the inter-session transfer scenario on both intact-limbed and amputee subjects. Let $\mathbf{M}$ be the polynomial expansion matrix of the training data, an optimal weight matrix $\mathbf{W}$ could be formulated as:
\begin{equation}
\begin{aligned}
\mathbf{W} = \underset{\mathbf{W}}{\mathrm{argmin}} \left\|\mathbf{M}\mathbf{W}-\mathbf{Y}\right\|^2.
\end{aligned}
\end{equation}
Similarly, the transfer learning process based on PC is defined as:
\begin{equation}
\begin{aligned}
\tilde{\mathbf{W}} = \sum_{i=1}^{i=K} \beta^{i}\mathbf{W}^{i} + \Bar{\mathbf{W}},
\end{aligned}
\end{equation}
where $\mathbf{W}^{i}$ and $\beta^{i}$ are the optimal weight matrix for the $i^{th}$ session and the corresponding weight ratio, $\Bar{\mathbf{W}}$ represents the optimal weight matrix on the new session and $\tilde{W}$ represents the adapted weight matrix. It is worth noticing that distance measurements such as Kullback–Leibler divergence~\cite{kullback1951information} could be used to select the source domain that's the most similar to the target domain to avoid negative transfer when there are multiple source domains available~\cite{kanoga2022subject}.

Next, we review main bio-inspired research endeavors under the linear assumption. As discussed in Section \ref{sec:emg_basics}, EMG signals are composed of superimposed MUAPTs generated from different MUs in both temporal and spatial domains. Muscle Synergy Modeling (MSM)~\cite{ajiboye2009muscle,muceli2013extracting,antuvan2016role, 4663628} has shown great success in terms of modeling the linear relationship between MUAPTs of muscles and the collected EMG signal. Let $x_m(t)$ be the generated MUAPTs from the $m^{th}$ muscle, define $\textit{act}_i(t) \in \mathbb{R}$ to be the activation signals, $x_m(t)$ could then be expressed as:
\begin{equation}
\begin{aligned}
x_m(t) = \sum_{i=1}^{i=N} g_{mi}\cdot \textit{act}_i(t),
\end{aligned}
\end{equation}
where $g_{mi}$ is the gain factor of muscle $m$ transferred to the $i^{th}$ activation signal with $N < M$. Assuming that only attenuation exists with distance but no filtering effect, the observed EMG signal at the $k^{th}$ electrode ($k^{th}$channel) is written as:
\begin{equation}
\begin{aligned}
y_k(t) &= \sum_{m=m}^{m=M}\sum_{i=1}^{i=N} l_{km} \cdot g_{mi} \cdot \textit{act}_i(t) \\
& = \sum_{i=1}^{i=N} a_{ki} \cdot \textit{act}_i(t), \\
\end{aligned}
\end{equation}
where $l_{km}$ is the factor that reflects the attenuation level from the $m^{th}$ muscle on the $k^{th}$ electrode and $a_{ki}$ is the combined weight factor that models both $l_{km}$ and $g_{mi}$. The above mixture could be written in matrix form:
\begin{equation}
\label{eq:nmf}
\begin{aligned}
Y(t) &= \mathbf{A} \cdot F(t),
\end{aligned}
\end{equation}
where $\textbf{A} \in \mathbb{R}^{K \times N}$ is the weighting matrix and $\textbf{F}$ is the synergy matrix. In EMG analysis, $Y$ is often observed, thus the solving for $\textbf{W}$ and $\textbf{F}$ becomes a linear blind source separation (BSS) problem~\cite{cichocki2009nonnegative}. Non-negative matrix factorization (NMF)~\cite{lee2000algorithms} finds an approximate solution to the equation \eqref{eq:nmf} with the constraint that all elements are non-negative.

Jiang \textit{et al.}~\cite{jiang2021data} proposed correlation-based data weighting (COR-W) for inter-subject transfer scenario of elbow torque modeling. In specific, they assume that the target domain data is a linear transformation of the source domain data, $X_{T} \approx \tilde{X}_{S} = \textbf{A}X_{S}$, where $\tilde{X}_{S}$ is the transformed source domain data. The underlying assumption is that the synergy matrix remains the same for both domains while the weighting matrix varies. A derived assumption of Jiang \textit{et al.} is that the covariance matrix of the transformed source domain should also be similar to the covariance matrix of the target domain data. The optimal matrix $A^*$ is estimated by minimizing the discrepancy between $\tilde{\Sigma_{S}}$ and $\Sigma_{T}$. The transformed source data is then used to re-train the model. Although Jiang \textit{et al.} proposed for inter-subject transfer scenario, while we argue that the linear assumption might not hold due to variation across subjects. Electrode shift, on the other hand, is reasonably more consistent with the linear assumption in practice. G\"{u}nay \textit{et al.}~\cite{gunay2019transfer} adopted MSM with NMF for knowledge transfer across different tasks. The weighting matrix $W$ calculated on the source domain is kept constant while the synergy matrix is re-estimated on the target domain data using the non-negative least squares (NNLS) algorithm.

In contrast to the works that map the source domain data to a new space, another line of work~\cite{prahm2017transfer,prahm2019counteracting, paassen2018expectation} transforms the target domain data so that the source domain objective prediction function is applicable again. Prahm \textit{et al.}~\cite{prahm2017transfer} viewed the target domain data as a disturbed version of the source domain data. The disturbance can be expressed as a linear transformation matrix $A$. The main aim is then to learn and apply an inverse disturbance matrix $A^{-1}$ to the target data such that the disturbance is removed. Prahm \textit{et al.}~\cite{prahm2017transfer} adopted Generalized Matrix Learning Vector Quantization (GMLVQ)~\cite{schneider2009adaptive} as the classifier and estimate the optimal $A^{-1}$ using gradient descent on the GMLVQ cost function. The linear transformation that maximizes the likelihood of disturbed data based on the undisturbed data could also be estimated by the Expectation and Maximization (EM) algorithm~\cite{moon1996expectation,paassen2018expectation}. Following their previous work~\cite{prahm2017transfer,paassen2018expectation}, Prahm \textit{et al.}~\cite{prahm2019counteracting} proposed that the linear transformation matrix could be further exploited based on the prior knowledge that the underlying EMG device is an armband with eight uniformly distributed channels. For the electrode shift scenario, Prahm \textit{et al.} assumed that the disturbed feature from channel $j$ could be linearly interpolated from neighboring channels from both directions with a mixing ratio $r$. Then the approximation of the linear transformation matrix is reduced to finding an optimal mixing ratio $r$.
\paragraph{Non-linear Transformation.}
The principle objective of feature transformation is to reduce the data distribution between the source and target domain. Thus, the metrics for measuring distribution difference is essential. Maximum Mean Discrepancy (MMD)~\cite{borgwardt2006integrating} is widely adopted in the field of transfer learning:
\begin{equation}
\begin{aligned}
\mathcal{MMD}(X_{T} ,X_{S}) &= \left\|\frac{1}{N^S}\sum_{i=1}^{i=N^S}\Phi(X_{S}^{i})-\frac{1}{N^T}\sum_{j=1}^{i=N^T}\Phi(X_{T}^{j})\right\|^2,
\end{aligned}
\end{equation}
where $\Phi$ indicates a non-linear mapping to the Reproducing Kernel Hilbert Space (RKHS)~\cite{berlinet2011reproducing}, $N^S$ and $N^T$ indicate the number of instances in the source and target domain, respectively. Essentially, MMD quantifies the distribution difference via calculating the distance between the mean vectors of the features in a RKHS. In addition to MMD, Kullback–Leibler divergence, Jenson–Shannon (JS) divergence~\cite{dagan1997similarity} and Wasserstein distance~\cite{shen2018wasserstein} are also common distance measurement criteria. 
\begin{figure*}[t]
\centerline{\includegraphics[width=\columnwidth]{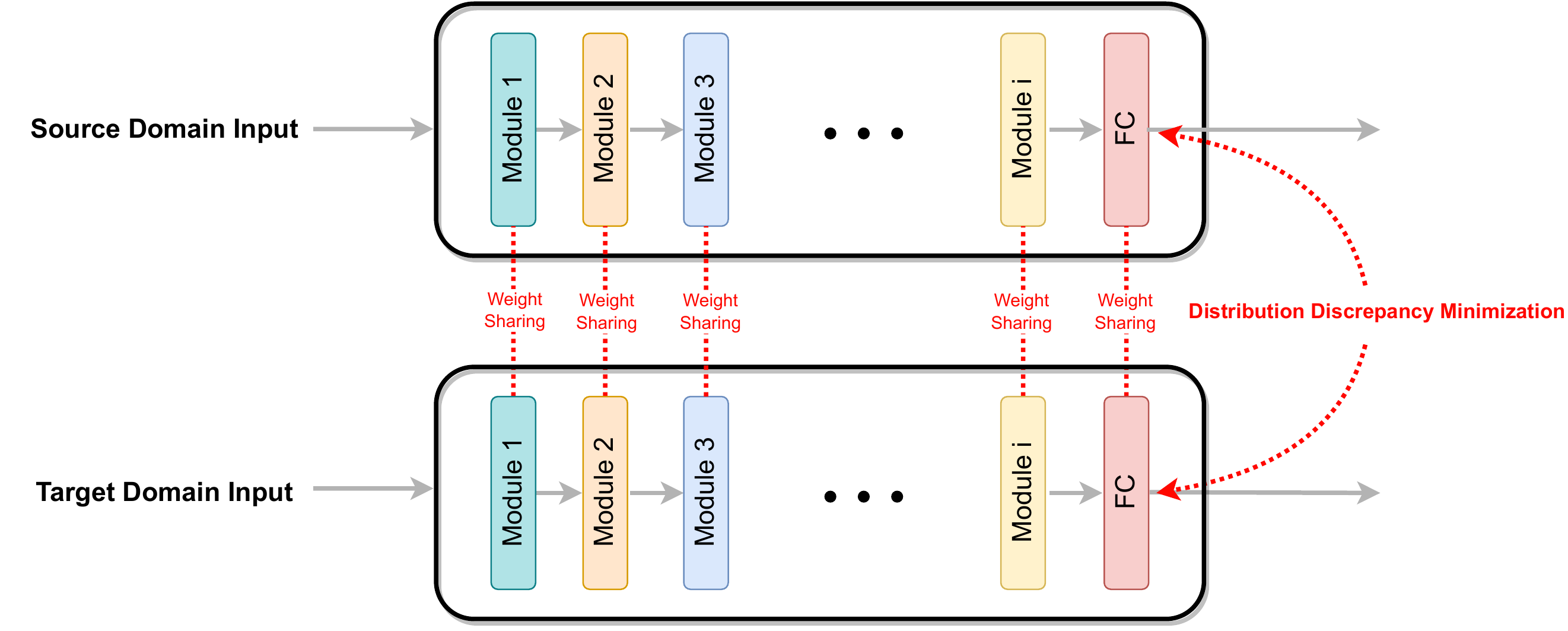}}
	\caption{
	Demonstration of applying the Siamese architecture for distribution discrepancy minimization. It is worth noticing that the design of neural network architectures varies across works. FC stands for the fully connected layer. The term 'module' refers to a combination of layers might contain convolution, normalization, or residual connection. The distribution discrepancy measurement is applied to the output of the FC layer just for demonstration. The distribution discrepancy measurement could essentially be applied to deep features output by any module.
	}
	\label{fig:siamese}
\end{figure*}
The Siamese architecture~\cite{chopra2005learning, wu2022bridging} is one commonly adopted architecture for DNN related transfer learning, as illustrated in Figure \ref{fig:siamese}. Zou \textit{et al.}~\cite{zou2021transfer} proposed a Convolutional Neural Network (CNN) based model named Multiscale Kernel Convolutional Neural Network (MKCNN) for hand gesture recognition. The authors proposed a transfer learning MKCNN (TL-MKCNN), which contains a Distribution Alignment Module (DAM) for inter-subject and inter-session transfer learning scenarios. TL-MKCNN adopts the Siamese architecture, with one network taking inputs from the source domain and the other one taking inputs from the target domain. The Siamese networks share weights with each other. DAM applies the JS divergence onto the output of the second layer of the fully connected layers to minimize the distribution difference between the deep features of the data from the source and target domain. Besides the Cross Entropy (CE) loss function for classification, Zou \textit{et al.} also apply a mean square error (MSE) to minimize the distance of instances to the corresponding class center. The overall loss function to train TL-MKCNN is the sum of JS divergence, CE, and MSE. Bao \textit{et al.}~\cite{bao2021inter} applied fast Fourier transform (FFT) to data segment and used the spectrum as input to their designed CNN based network. Similar to ~\cite{zou2021transfer}, the MMD loss is applied to the output of the second fully connected layer. A Regression Contrastive Loss is proposed to minimize the distance in the feature space between the source domain instance and target domain instance of the same category. Normalization tricks are adopted to modify the loss for regression tasks.

\begin{figure*}[t]
\centerline{\includegraphics[width=\columnwidth]{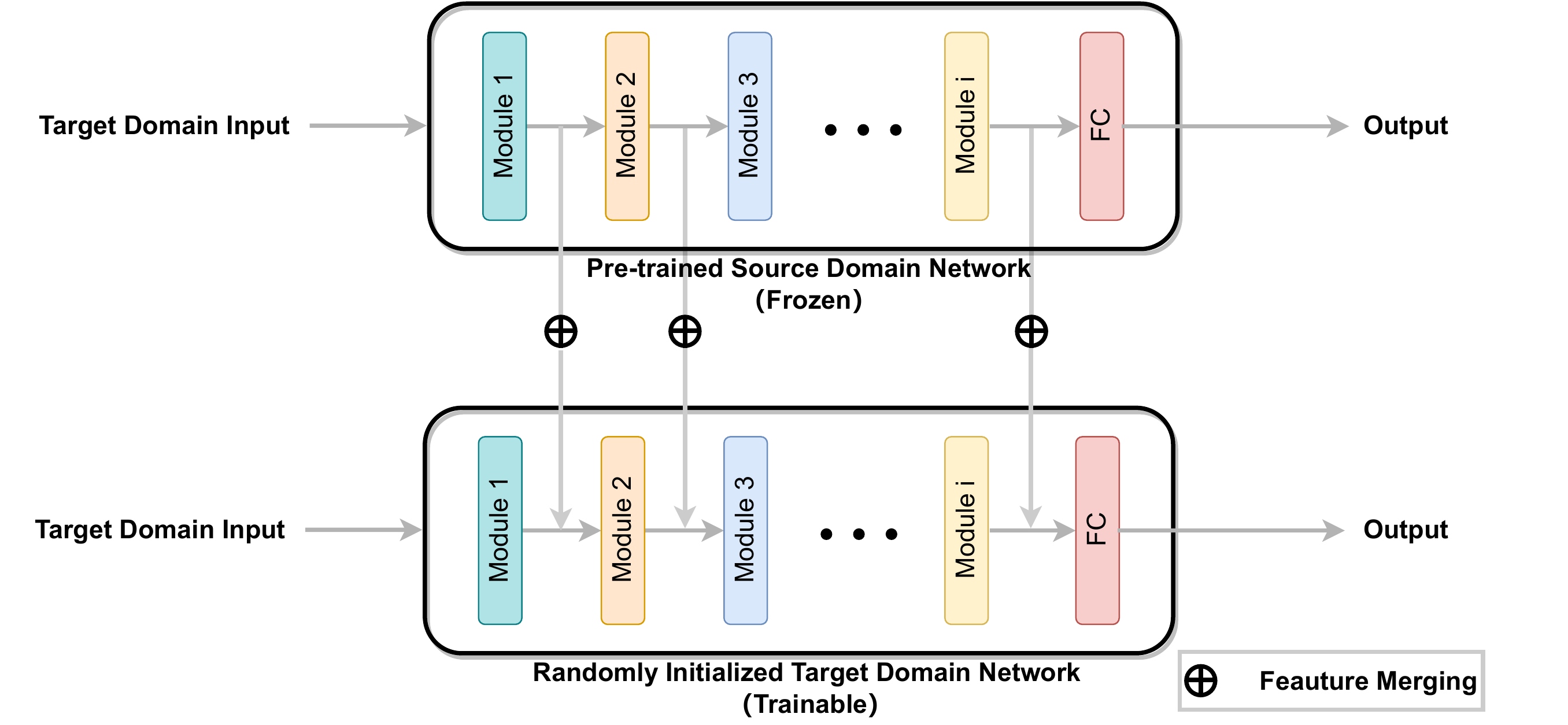}}
	\caption{
	Illustration of the architecture of the progressive neural network. Frozen indicates that the parameters of the network are fixed while trainable suggests that the network parameters will be updated during training. The same input is fed to both networks, the intermediate features from each module of the pre-trained network is merged with corresponding intermediate features of the target domain network.
	}
	\label{fig:pnn}
\end{figure*}

C\^{o}t\'{e}-Allard \textit{et al.}~\cite{cote2019deep,cote2017transfer} proposed to use the Progressive Neural Network (PNN)~\cite{rusu2016progressive} to alleviate \textit{catastrophic forgetting} caused by directly fine-tuning the network parameters with data from the target domain. As shown in Figure \ref{fig:pnn}, a source domain network is first trained with data from the source domain. The model parameters of the source domain network are then fixed while the parameters for the target domain network is randomly initialized. Note that the network structures of both networks are exactly the same except for the model parameters. During the transfer learning process, target domain instances are fed to both networks. The intermediate features of each module of the source domain network is then merged with the corresponding features of the target domain network and fed forward to the next module of the target domain network. The underlying hypothesis is that although distribution variation exists between the source and target domain, generic and robust features could be attracted for more effective representation learning.

Du \textit{et al.}~\cite{du2017surface} proposed to adopt Adaptive Batch Normalization (AdaBN)~\cite{li2016revisiting} for inter-session transfer learning. AdaBN is a lightweight transfer learning approach for DNNs based on Batch Normalization (BN)~\cite{ioffe2015batch}. BN was initially proposed to accelerate the convergence of the DNN for faster CNN training. Formally, define $\mathbf{Z}=[\mathbf{z}_i]_{i=1}^{B}$ to be a batch of intermediate features of instances with batch size $B$, the BN layer transforms $\mathbf{Z}$ as follows:
\begin{equation}
\begin{aligned}
\tilde{\mathbf{z}} = \gamma \cdot \frac{\mathbf{z}_j - \mathbb{E}[\mathbf{Z}._{j}]}{\sqrt{Var[\mathbf{Z}._{j}]}} + \beta,
\end{aligned}
\end{equation}
where $\gamma$ and $\beta$ are learnable parameters, $Var$ stands for variance. The underlying hypothesis is that labeled related knowledge is stored in the network parameters of each layer, and the domain related knowledge is portrayed by the statistics of the BN layers. The transformation ensures that the distribution of each layer remains the same over mini-batches so that each layer of the network receive input of similar distribution regardless of the source or target domain. Different from fine-tuning, AdaBN doesn't require target domain label for knowledge transfer and only a small fraction of the network parameters need to be updated. In particular, the network is first pre-trained on source domain data. During the training process, the statistics of BN layers are calculated by applying a moving average for all data batches. All network parameters are fixed except for the parameters of BN layers during transfer learning. The update of BN statistics to target domain data could easily be done by a forward pass.
\subsection{Model Based Perspective}
From the model perspective, transfer learning approaches can also be interpreted in terms of model parameters and model structures.
\begin{figure*}[t]
\centerline{\includegraphics[width=0.8\columnwidth]{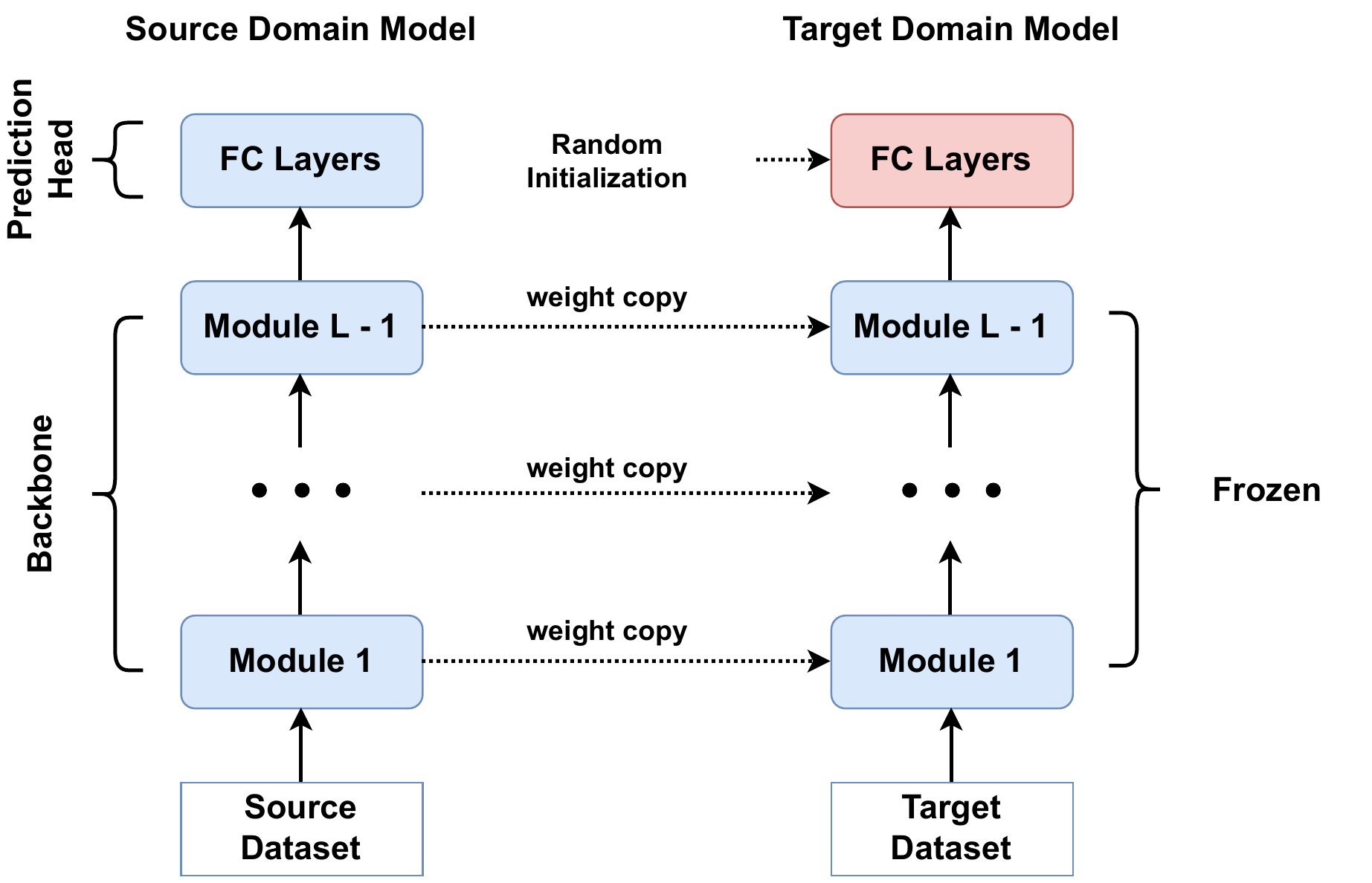}}
	\caption{
	Illustration of transferring knowledge by sharing the weights of the neural network. The weights of the backbone modules are first copied to the target domain network and frozen. The term 'module' refers to a combination of layers that might contain convolution, normalization, or residual connection. FC stands for the fully connected layer. The weights of the prediction head are randomly initialized and trained from scratch.
	}
	\label{fig:param_sharing}
\end{figure*}
\subsubsection{Parameter Fine-tuning}
\label{sec:finetune}
One intuitive way of transferring knowledge of DNN is to tune the network parameters of the source learner using data from the target domain. Fine-tuning~\cite{donahue2014decaf} refers to the training process where the network is first trained on one dataset (large-scale) and use the network parameters as initialization to further train on another dataset (small scale). Fine-tuning is a common strategy in the Computer Vision (CV) community where the neural networks are first pre-trained on ImageNet (IN) either in a supervised manner or self-supervised manner and later fine-tuned for various downstream tasks such as classification~\cite{wu2021align} and object detection~\cite{li2022architecture}. IN-21K is a large scale dataset with 15 million images over 2200 classes. The underlying assumption is that the dataset for downstream tasks (target domain) is of similar data distribution of IN (source domain). Extensive experiments have shown that using IN pre-trained weights and fine-tuning improves performance by a large margin. Inspired by the success of fine-tuning in CV, authors~\cite{rezaee2022hybrid,bird2020cross,demir2019surface} first transform EMG signal to 2D data via Short-time Fourier Transform (STFT) and treat the spectrogram as image input to the neural network. It is worth noticing that these works~\cite{rezaee2022hybrid,bird2020cross,demir2019surface} use IN pre-trained network to improve the performance of the network that takes STFT as input. This line of work do not fall into any of the transfer learning scenarios that we summarized in Section \ref{sec:trans_setting}. The neural network learns feature extraction ability (texture and shape) and the ability to localize to the foreground object. IN mainly contains images of natural scenes such as objects, animals, and humans. Since the gap between the source domain (natural scenes) and the target domain (spectrum image) is tremendous, it is questionable as to what knowledge is transferable. Phoo \textit{et al.}~\cite{phoo2020self} compared the transfer performance of using miniIN (a small subset of IN) as source domain and using IN as source domain to ChestX (X-ray images for chest)~\cite{wang2017chestx} as target domain. Experimental results show that pre-training on IN yields no better performance than on miniIN and both yields poor diagnosis accuracy. This suggests that more data does not help improve the generalization ability, given that no more informative knowledge can be extracted from the source domain to benefit the target domain learner. Pre-training the network on the source domain and then using the pre-trained weights to initialize the neural network for further training using the target domain data is another popular fine-tuning strategy for EMG transfer learning~\cite{lehmler2021deep, ameri2019deep, kobylarz2020thumbs, dao2019deep,tam2021intuitive}. There would be little constraint nor assumption on the transfer scenarios since this transfer process is simple and can be viewed as sequentially train the network with two datasets. When there are EMG data recorded from multiple subjects or sessions, it is possible to combine the data and treat the combined data as the source domain~\cite{sloboda2021utility,zakia2021force}. Or it is also a solution to train a unique model for each subject or session and to select a certain number of models that give the best performance on the target domain~\cite{kim2019subject,hoshino2022comparing}, the selected models are then fine-tuned on the target dataset to provide final prediction based on majority voting~\cite{penrose1946elementary}. However, fine-tuning suffers from the \textit{catastrophic forgetting}, meaning that knowledge from the source domain will be forgotten by the neural network rapidly upon the introduction of target domain data~\cite{french1991using}. Besides the parameters fine-tuning of DNNs, the parameters of Decision Trees~\cite{quinlan1986induction} (DTs) could also be fine-tuned for EMG transfer learning~\cite{marano2021questioning}. The motivation is that the structure of decision trees for similar tasks should be similar and the domain difference is reflected from different decision threshold values associated with the features. Structure Transfer (STRUT)~\cite{segev2016learn} first discards all the numeric threshold values of learned trees on the source domain data and selects a new threshold value $\tau (\nu)$ for a node $\nu$ given that the subset of target examples reach $\nu$ in a top-down manner. Any node $\nu$ that's empty in terms of target domain data is considered unreachable and will be pruned. Define $\tau$ to be the threshold value of feature $\phi$ at node $\nu$ that splits any set of labeled data $S_{\nu}$ into two subsets, denoted $S_{L}$ and $S_{R}$. $P_L$ and $P_R$ denote the label distribution of $S_{L}$ and $S_{R}$, respectively. STRUT aims to find a new threshold $\tau^{'}$ with maximum Divergence Gain (DG) subject to the condition where the new thresholds are local maximums of Information Gain (IG)~\cite{quinlan1986induction}:
\begin{equation}
\begin{aligned}
\mathcal{DG} &= 1 - \frac{\left\|S_{L}^{T}\right\|}{\left\|S_{\nu}^{T}\right\|}\mathcal{JSD}(Q_{L}^{T},Q_{L}^{S}) -  \frac{\left\|S_{R}^{T}\right\|}{\left\|S_{\nu}^{T}\right\|}\mathcal{JSD}(Q_{R}^{T},Q_{R}^{S}),
\end{aligned}
\end{equation}
where $\left\| \cdot \right\|$ stands for the cardinality, $S$ and $T$ on the superscript stand for the source and target, respectively.

\subsubsection{Parameter Sharing}
The neural network architectures are not specified in Section \ref{sec:finetune} since parameter fine-tuning tunes all parameters of the network regardless of various network designs. It is stated that fine-tuning the whole network suffers from \textit{catastrophic forgetting} and knowledge learned from the source domain will be quickly forgotten. In most of the works~\cite{lehmler2021deep, ameri2019deep, kobylarz2020thumbs, dao2019deep,tam2021intuitive} that adopt fine-tuning, the target domain dataset is of the same size as the source domain dataset. Consider the case where the target domain dataset is small compared to the source domain, with forgotten knowledge from the source domain, the neural network is prone to suffer from over-fitting~\cite{hawkins2004problem}. A possible solution is to freeze partial network parameters and to only update partial parameters during the fine-tuning process. An illustration of knowledge transferring via parameter sharing is provided in Figure \ref{fig:param_sharing}. A neural network design could be roughly divided into the backbone and the prediction head. The backbone serves as the feature extractor and is usually CNN based or Recurrent Neural Networks (RNN) based. The prediction head is usually composed of fully connected layers and predicts the desired labels based on the deep features extracted by the backbone. Assuming that the extracted deep features are generic for various transfer scenarios, the weight of the backbone could be frozen once pre-trained on the source domain dataset to prevent \textit{catastrophic forgetting}~\cite{chen2020hand,ketyko2019domain,fan2021improving,soselia2022lower,yu2021surface,tsinganos2021transfer,li2021transfer}. Only the fully connected layers of the prediction head need to be updated which reduces transfer training time and guarantees fast convergence. 

\subsubsection{Model Structure Calibration}
Besides knowledge transferring via trained parameters, next we explore the possibility of EMG transfer learning from the model structure perspective. Since it is often the case that there is a lack of labeled data in the target domain and as such it might not be sufficient to construct a reliable high performance model solely on the target domain data, optimizing the model structure of a pre-trained model to fit the target domain data is desired. As we mentioned in the previous section that DNNs are believed to be able to extract generic features, thus it is impractical and time consuming to alter or even search for neural network structures using Neural Architecture Search (NAS)~\cite{elsken2019neural} for various domains. However, Random Forest (RF)~\cite{breiman2001random} on the other hand, is more suitable for structure calibration since knowledge transfer could be done by pruning or growing the source tree model. Marano \textit{et al.}~\cite{marano2021questioning} proposed to use structure expansion/reduction (SER)~\cite{segev2016learn} for EMG based hand prostheses control. As the name suggests, the SER algorithm contains two phases: \textit{expansion} and \textit{reduction}. Consider an initial random forest that is induced using the source domain 
data. In the \textit{expansion} phase, SER first calculates all labeled data points in the target domain dataset that reaches node $\nu$ and then extends node $\nu$ into a full tree. In the \textit{reduction} phase is performed to reduce the model structure in a bottom-up fashion. Define $\mathcal{E}_{sub}$ to be the empirical error of the subtree
with root node $\nu$, $\mathcal{E}_{leaf}$ denotes the empirical error on node $\nu$ if $\nu $ were to be pruned to a leaf node. The subtree is to be pruned into a node leaf if $\mathcal{E}_{sub} > \mathcal{E}_{leaf}$. SER is performed on each decision tree separately and the resulting random forest is the adapted model for the target domain data.
\subsubsection{Model Ensemble}
Combining data from various sources into a single source domain may not yield satisfactory results since the distributions of these domains might vary greatly from each other. Another commonly adopted strategy for EMG transfer learning is model ensemble. The model ensemble aims to combine a set of weak learners to make the final prediction. Some previously reviewed EMG transfer learning approaches already adopted this strategy. For instance, Kim \textit{et al.}~\cite{kim2019subject} proposed to train a unique classifier for each subject and further fine-tune the top ten best performing classifiers on a new target subject. The final prediction is the most commonly classified by the ensemble of all ten fine-tuned classifiers. Decision Trees are another popular choice for weak learners. Zhang \textit{et al.}~\cite{zhang2021feature} proposed feature incremental and decremental learning method (FIDE) based on Stratified Random Forest (SRF) for knowledge transfer with missing or added electrodes. In specific, define $S_i$ and $S_j$ to be the electrode sketch score~\cite{liberty2013simple} for electrode $e_i$ and $e_j$, respectively. The distribution difference between electrodes $e_i$ and $e_j$ is defined as:
\begin{equation}
\begin{aligned}
\mathcal{DD}(i,j) &= \frac{\rho(S_i,S_j) + \psi(e_i.e_j) + 1}{4},
\end{aligned}
\end{equation}
where $\rho(\cdot)$ stands for the Pearson Correlation Coefficients (PCC) and $\psi$ denotes the inverse of the Euclidean distance between $e_i$ and $e_j$. K-means~\cite{macqueen1967classification} is then utilized to cluster the electrodes into $K$ clusters based on the DD. Denote $M$ as the number of weak learners in the ensemble model, SRF is built on the source domain data where $\lceil M/K \rceil$ trees are induced using data collected with electrodes in the corresponding cluster. If electrode $i$ is missing in the target domain data, the missing features could be recovered from the most similar electrode $j$. If there are incremental electrodes in the target domain dataset, FIDE first selects set of weak learners to be updated based on a performance score:
\begin{equation}
\begin{aligned}
\mathcal{S}(m) &= acc(h_m) + \frac{\# feature_m}{\# feature},
\end{aligned}
\end{equation}
where $h_m$ stands for the $m^{th}$ decision tree, $\# feature_m$ denotes the number of features used by $h_m$, and ${\# feature}$ denotes the total number of features. Top $M*\delta$ weak learners are then selected for updated where $\delta \in [0,1]$. The SER and STRUT algorithms~\cite{segev2016learn} introduced in previous sections are again used for transfer learning on decision trees. Compared to the majority voting way of ensemble, FIDE updates the source domain model to extract new knowledge from target domain data while not abandoning the already learned knowledge.

\begin{algorithm}[t]
    \caption{MAML Style Meta-learning for Transfer Learning}
    \SetAlgoLined
    \textbf{ Input }: 

    Task distribution : $p(\mathcal{T})$,
    Loss function : $\mathcal{L}$,
    learning rate for inner loop: $\alpha$,
    learning rate for outer loop: $\beta$ \\
    \textbf{ Output }: 
    Prediction Model : $f_\Theta$, \\
    \textbf{ Initialization }:Randomly initialize $\Theta$\\
    \While{not done}
    {
        {Sample a batch of tasks $\mathcal{T}_i$ from $p(\mathcal{T})$}\\ 
    \For{all task$\mathcal{T}_i$}{
    Evaluate error $\mathcal{L}_{\mathcal{T}_i}(f_{\Theta})$ with respect to the $D_j^{train}$ \\
    Update $\Theta$ with gradient descent: \\
    $\Tilde{\Theta} \longleftarrow \Theta - \alpha \cdot \frac{ \partial \mathcal{L}_{\mathcal{T}_i}(f_{\Theta} }{\partial \Theta}$\\
    % , D_j^{test}
    }
    Evaluate error $\mathcal{L}_{\mathcal{T}_i}(f_{\Tilde{\Theta}})$ with respect to the $D_j^{test}$ \\
    Update $\Theta$ with gradient descent: \\
    $\hat{\Theta} \longleftarrow \Theta - \beta \cdot \frac{ \partial \mathcal{L}_{\mathcal{T}_i}(f_{\Tilde{\Theta}}) }{\partial \Theta}$\\

    }

\label{alg:1}
\end{algorithm}

\subsection{Training-scheme Based Perspective}
In addition to the previously mentioned approaches that can be subsumed into pre-defined paradigms, we also review works that design special training schemes for EMG transfer learning. Zhai \textit{et al.}~\cite{zhai2017self} proposed a self re-calibration approach for inter-session hand prosthesis control. In particular, a source domain classifier is first trained with EMG data of existing sessions. Given the target domain data, each EMG data segment $x^i$ is assigned a prediction label $y^i$ by applying a forward pass of the EMG segments. Based on the assumption that temporally adjacent EMG segments are likely to be generated from the same hand movement, the assigned labels are re-calibrated with majority voting:
\begin{equation}
\begin{aligned}
\tilde{y}^i \longleftarrow \textit{Majority Voting}(f_{\mathcal{S}}(x^{i-k}, x^{i-k+1},\dots,x^{i},\dots,x^{i+k})),
\end{aligned}
\end{equation}
where $f_{\mathcal{S}}$ is the source domain classifier and $k$ indicates the number of neighboring segments used to re-calibrate the label from both directions in time before and after $x^i$. Then the target domain data with re-calibrated labels are used to update the source domain classifier. It is worth noticing that such a transfer scheme does not require target domain data and can be easily adopted for day-to-day re-calibration.

Meta-learning~\cite{hospedales2020meta} is another training paradigm that can be used for EMG transfer learning. Meta-learning is commonly known as learning to learn~\cite{thrun1998learning}. In contrast to conventional machine learning algorithms that optimize the model over one learning episodes, meta-learning improves the model over multiple learning episodes. The meta-learning goal of generalizing the model to a new task of an incoming learning episode with limited samples aligns well with the notion of transfer learning. Intuitively speaking, meta-learning divide the source domain data into multiple learning episodes, with each containing a few samples and mimicking the transfer processing during training so that the model trained has good transferability in terms of the true target domain. Rahimian \textit{et al.}~\cite{rahimian2021few} proposed meta-learning based training scheme called Few-Shot Hand Gesture Recognition (FHGR) for the transfer case where only a minimal amount of target domain data are available for re-calibration. Define a N-way k-shot few shot learning problem, let $\mathcal{T}_j= \{ D_j^{train}, D_j^{test}, \mathcal{L} \}$ denote a task associated with the source domain dataset where $D_j^{train}= \{ (x_i,y_i) \}_{i=1}^{K \times N}$ and $\mathcal{L}$ is a loss function to measure the error between the prediction and the ground-truth label. Please be aware that the task $\mathcal{T}$ here is a naming convention in the meta-learning area and is of a different meaning than the task that we define for a domain. FHGR aims to predict the labels of $ D_j^{test}$ based on the samples seen from $D_j^{train}$ consisting of $K$ samples from each of the $N$ classes over a set of tasks samples from $p(\mathcal{T})$. A Pseudocode in the MAML style~\cite{finn2017model} is provided in Algorithm \ref{alg:1}.

EMG transfer learning could also benefit from data augmentation via generating synthetic data as data from other sessions or subjects (target domain data). Generative Adversarial Networks (GANs) are a famous type of networks for data generation without explicitly modeling the data probability distribution. A typical GAN contains a generator $G$ and the discriminator $D$ which are two neural networks. A random noise vector sampled from a Gaussian or uniform distribution is input to the generator network to produce a sample $x_g$ that should be similar to a real data sample $x_r$ drawn from a true data distribution $P_r$. Either $x_r$ or $x_g$ is input to the discriminator to get a classification result of whether the input in real or fake. Intuitively, the generator aims to generate fake samples that could confuse the discriminator as much as possible, while the task of the discriminator is to best distinguish fake samples from real ones. The training objective of GAN can be defined as:
\begin{subequations}
\begin{align}
\mathcal{L}_D &= \underset{D}{max}\mathbb{E}_{x_r}  [logD(x_r)]  + \mathbb{E}_{x_g}[log(1-D(x_g))] \\
\mathcal{L}_G &= \underset{D}{max}\mathbb{E}_{x_g}[log(1-D(x_g))]
\end{align}
\end{subequations}
 Zanini \textit{et al.}~\cite{zanini2020parkinson} adopted DCGAN~\cite{radford2015unsupervised} which is an convolution-based extension of the very original GAN and style transfer for Parkinson’s Disease EMG data augmentation.
 
 Besides GANs, style transfer has also been utilized to augment EMG data. Given a piece of fine art work, painting, for example, humans have the ability to appreciate the interaction of content and style. "The Starry Night" by Van Gogh is an appealing painting that attracts a lot of re-drawing attention which follows the same drawing style of Van Gogh but with different content. Gatys \textit{et al.}~\cite{gatys2015neural} proposed an algorithm for artistic style transfer that combines content from one painting and the style of another painting. A similar idea could be extended to EMG signals for transfer learning. An EMG signal can also be regarded as the interaction of content and style. The style might refer to the biological characteristics of the subject, such as muscle condition, the filtering effect of a recording device, or simply a session. The content depicts the spikes carrying moving intention from the neural system to the corresponding muscles. Consider that the content of the different muscle movement are the same regardless any other conditions, the style component then process the control signals for moving to subject, device, or session specific data.  Zanini \textit{et al.}~\cite{zanini2020parkinson} adopted 
style transfer~\cite{gatys2015neural} to augment Parkinson’s Disease EMG data of different patterns. Specifically, given a content EMG signal $e_c$ and a style image $e_s$, the algorithm aims to find an EMG signal $e$ that's of the same content as $e_c$ and of the same style as $e_s$. Mathematically, the transferring process minimizes the following loss function:
 \begin{subequations}
\begin{align}
% e^* &= \underset{e}{argmin}\mathcal{L}(e,e_c,e_s) \\
% &= \underset{e}{argmin} \alpha*\mathcal{L}_c(e,e_c) + \beta * \mathcal{L}_s(e,e_s)\\
\mathcal{L}_c(e,e_c) &= \sum_{l} \left\| \mathcal{F}^{l}(e_c) - \mathcal{F}^{l}(e)  \right\|^2 \\
\mathcal{L}_s(e,e_s) &= \sum_{l} \left\| \mathcal{G}(\mathcal{F}^{l}(e_c)) - \mathcal{G}(\mathcal{F}^{l}(e))  \right\|^2,
\end{align}
\end{subequations}
where $\mathcal{F}(\cdot)$ is the output feature of the $l^{th}$ layer of the neural network, $\mathcal{G}$ stands for the Gram matrix~\cite{drineas2005nystrom}. The content component and style component are controlled by two hyper-parameters.
 \begin{equation}
\begin{aligned}
\mathcal{L} &= \alpha*\mathcal{L}_c + \beta * \mathcal{L}_s\\
\end{aligned}
\end{equation}

 Besides directly generating EMG data, Suri \textit{et al.}~\cite{suri2018transfer} proposed to synthesize extracted features of EMG signals with an LSTM network\cite{gers2000learning} to mimic EMG data from other subjects or different sessions. Different from GAN and style transfer based EMG augmentation that are directed by loss functions that either measure the authenticity or similarity, the method proposed by Suri \textit{et al.} simply relies on the assumption that extracted features are robust and that EMG signal generated by altering features are correlated to the recorded real data.  
\subsection{Adversarial Based Perspective}
\begin{figure*}[t]
\centerline{\includegraphics[width=1.1\textwidth]{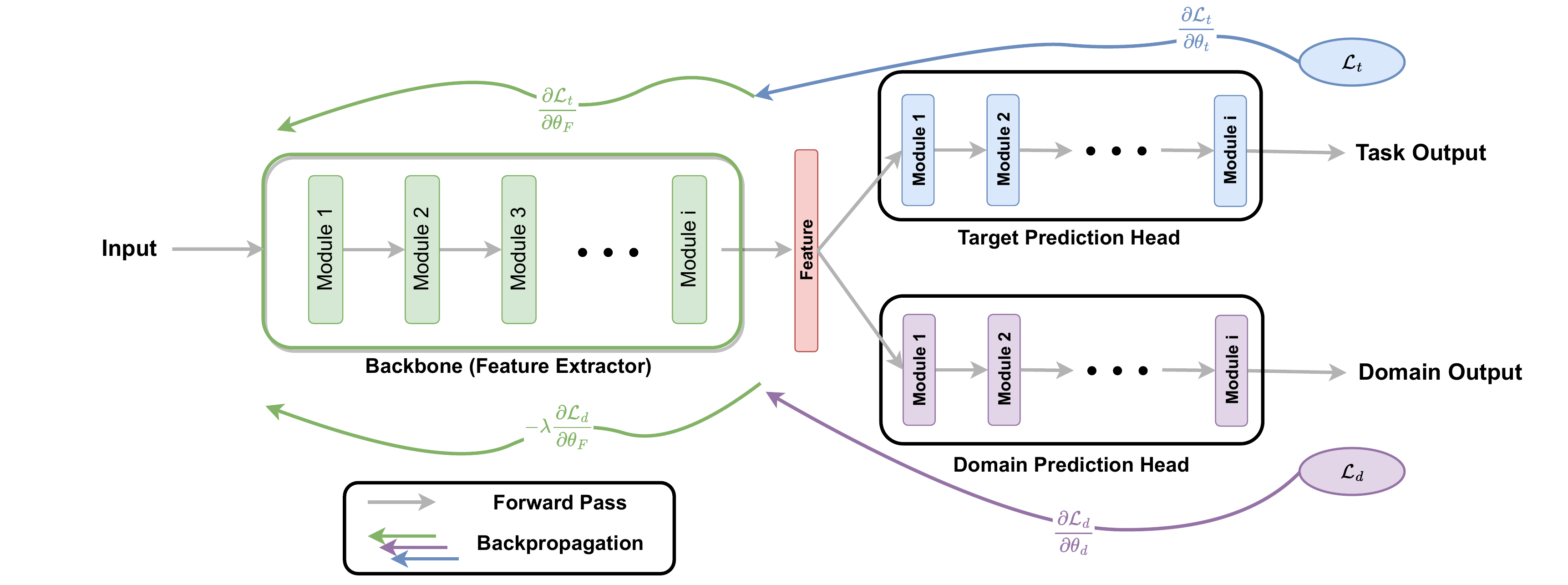}}
	\caption{
	Illustration of a typical DANN. A backbone of any arbitrary design for feature extraction is marked in green while the task prediction head and domain prediction head are marked in blue and purple, respectively. The output deep feature from the backbone is fed to both heads for loss calculation with respect to the ground truth label. The gradient of $\mathcal{L}_t$ is backpropagated through the task prediction head and the backbone for parameter update. The domain prediction head is updated by the gradient of $\mathcal{L}_d$. The negative gradient from $\mathcal{L}_d$ also flows back to the backbone for parameter update. 
	}
	\label{fig:DANN}
\end{figure*}

Recall that in Section \ref{sec:feature}, we introduce non-linear feature based approaches that reduce the data distribution by explicit deep feature transformation. In this section, we review a set of methods that force the neural network to learn hidden EMG representations that contain no discriminative information in terms of the origin of the data for domain generic feature extraction. With this objective, Domain-Adversarial Neural Networks (DANN)~\cite{ganin2016domain} is a type of neural network that contains a backbone $\mathcal{F}(\cdot; \theta_{\mathcal{F}})$ parameterized by $\theta_{\mathcal{F}}$ for feature extraction and two prediction heads: one for predicting the task label and another for predicting the origin of the data (source or target domain). We refer to the prediction head for the source domain task as the task prediction head $\mathcal{P}_{t}(\cdot; \theta_t)$ and refer to the prediction head for domain classification as domain prediction head $\mathcal{P}_{d}(\cdot; \theta_d)$. The parameters of the network are optimized in a way that the learned deep feature minimizes the loss for the task prediction head while maximizing the loss for the domain prediction head. The domain prediction head works adversarially to the task prediction head hence the name DANN. Formally, the overall loss function for optimizing $\theta_{\mathcal{F}}$, $\theta_t$ and $\theta_d$ is defined as:
 \begin{equation}
\begin{aligned}
\label{eq:dann}
\mathcal{E}(\theta_{\mathcal{F}}, \theta_t, \theta_d) &= \frac{1}{n}\sum_{i=1}^{n} \mathcal{L}_{t}(\theta_t, \theta_{\mathcal{F}})^{i} - \lambda(\frac{1}{n}\sum_{i=1}^{n} \mathcal{L}_{d}(\theta_d, \theta_{\mathcal{F}})^{i} + \frac{1}{m}\sum_{j=1}^{m} \mathcal{L}_{d}(\theta_d, \theta_{\mathcal{F}})^{j}),
\end{aligned}
\end{equation}
where $\mathcal{L}_{t}$ denotes the loss function for the source domain prediction task,  $\mathcal{L}_{d}$ denotes the loss function for the domain classification, $\lambda$ is a balance factor, $n$ and $m$ indicate the number of the source domain data and target domain data, respectively. The parameters $\theta_{\mathcal{F}}$, $\theta_t$ and $\theta_d$ and then are updated using gradient descent:
 \begin{equation}
\begin{aligned}
\theta_{\mathcal{F}} &\longleftarrow \theta_{\mathcal{F}} - \beta (\frac{ \partial \mathcal{L}_{t} }{\partial \theta_{\mathcal{F}}} - \lambda (\frac{ \partial \mathcal{L}_{d} }{\partial \theta_{\mathcal{F}}})), \\
\theta_t &\longleftarrow \theta_t - \beta \frac{ \partial \mathcal{L}_{t} }{\partial \theta_t}, \\
\theta_d &\longleftarrow \theta_d - \beta \lambda \frac{ \partial \mathcal{L}_{d} }{\partial \theta_d},
\end{aligned}
\end{equation}
where $\beta$ is the learning rate. We provide an illustration of data and gradient flow of DANN in Figure \ref{fig:DANN}.

C\^{o}t\'{e}-Allard \textit{et al.}~\cite{cote2021transferable} proposed to use DANN for multi-domain for inter-session EMG transfer learning. During training, each mini-batch contains randomly sampled EMG segments from one session. Each mini-batch is assigned with a class index indicating different sessions for the domain predicting labels. A gradient reversal layer~\cite{ganin2016domain} is adopted for easy implementation of negative gradient flow from the domain prediction loss to the backbone. Note that the task prediction head is only updated with loss from the source domain data. In a contemporaneous work, C\^{o}t\'{e}-Allard \textit{et al.}~\cite{cote2020unsupervised} also explored using Virtual Adversarial Domain Adaptation (VADA)~\cite{shu2018dirt} together with Decision-boundary Iterative Refinement Training with a Teacher (DIRT-T)~\cite{shu2018dirt} for adversarial based EMG transfer learning. VADA is an extension of DANN that incorporates locally-Lipschitz constraint via Virtual Adversarial Training (VAT)~\cite{miyato2018virtual} to punish the violation of the cluster assumption during training. On top of the trained model by VADA, DIRT-T aims to optimize the decision boundary on the target domain data by fine-tuning the model. In specific, the model parameter from the previous iteration is treated as the teacher model, the optimization goal is to seek a student model that is close to the teacher model while minimizing the cluster assumption violation. Following the work of C\^{o}t\'{e}-Allard~\textit{et al.}, other DANN related EMG transfer learning research endeavors~\cite{sohn2022feasibility,campbell2021deep} were made for various transfer scenarios.
\begin{figure*}[t]
\centerline{\includegraphics[width=\columnwidth]{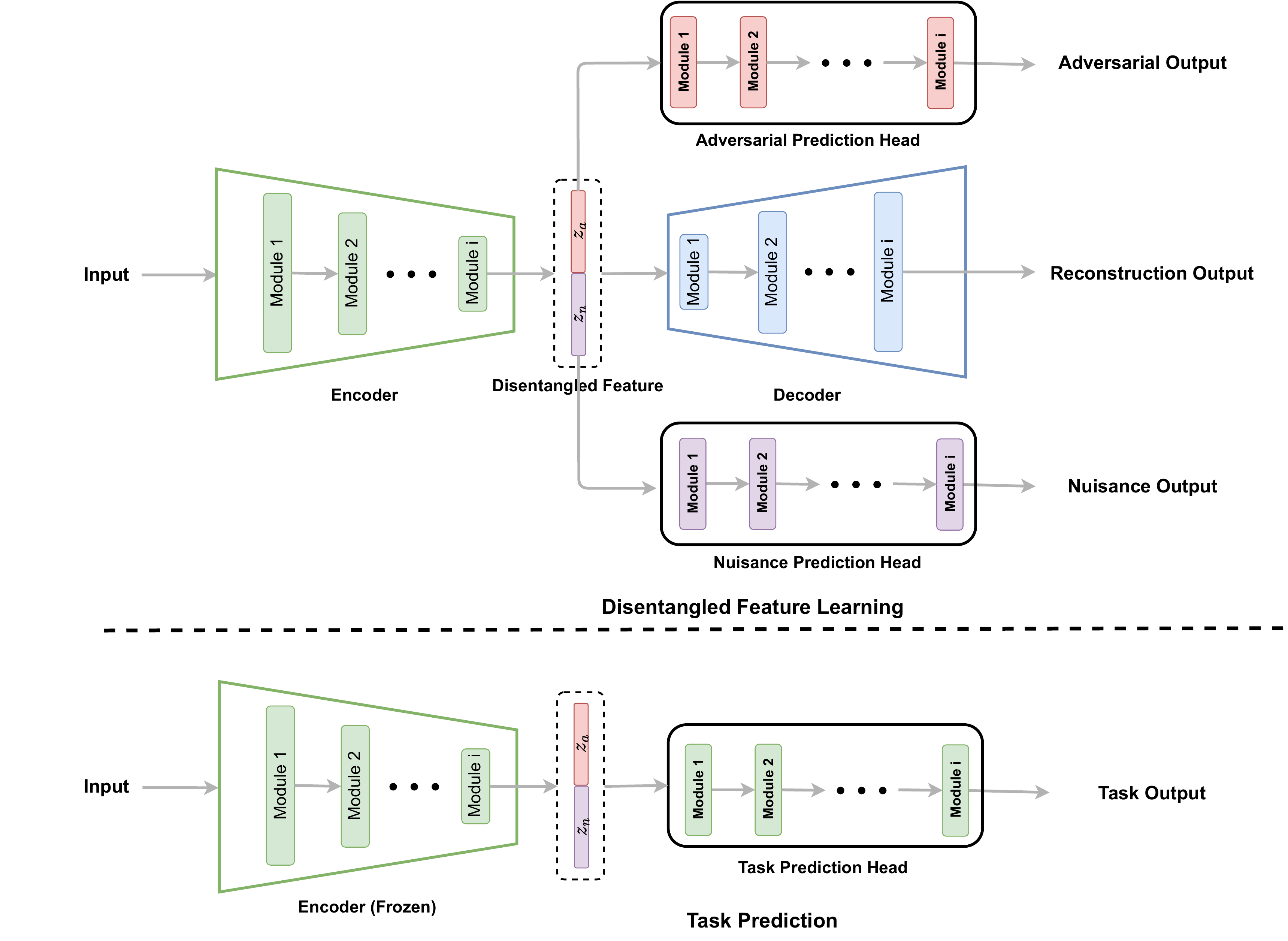}}
	\caption{
	Illustration of Disentangled Adversarial Autoencoder (DAA). The disentangled feature learning phase is demonstrated above the dotted line, while the task prediction phase is shown below the dotted line. In the disentangled feature learning phase, the input data is mapped into disentangled feature representation $z_a$ and $z_n$ with each passed to the corresponding prediction head. The overall latent representation is passed to the decoder for signal reconstruction. After feature learning, two prediction heads with the decoder are discarded. A new task prediction head with random weights is introduced on top of the encoder with frozen weight for task prediction.
	}
	\label{fig:daa}
\end{figure*}
Han \textit{et al.}~\cite{han2020disentangled} further proposed Disentangled Adversarial Autoencoder (DAA) which disentangles the learned latent representation into adversary and nuisance blocks to model task-related features and domain-related features disjointly. Based on the autoencoder (AE)~\cite{hinton1993autoencoders} structure, the encoder $\mathcal{F}(\cdot; \theta)$ maps the input signal $x$ into a latent representation $z=[z_a, z_n]$ where $z_a$ and $z_n$ stand for the adversary and the nuisance sub-representation, respectively. $z_a$ is expected to contain only the task relevant feature but no domain-specific information $i_d$. On the other hand, the encoder embeds sufficient domain-specific data into $z_n$. The decoder $\mathcal{G}(\cdot; \eta)$ reconstructs the original input signal based on latent representation $z$. Similar to DANN, DAA also adopts two prediction head: adversarial prediction head $\mathcal{P}_a(\cdot; \phi)$ and nuisance prediction head $\mathcal{P}_n(\cdot; \psi)$. Formally, the overall loss to train DAA is defined as: 
 \begin{equation}
\begin{aligned}
\label{eq:daa}
\mathcal{L}(\theta, \phi, \psi, \eta) &= - \lambda_n \mathbb{E}[\log p(i_d|z_n)] + \lambda_a \mathbb{E}[\log p(i_d|z_a)] + \mathbb{E}[\left\| x - \mathcal{G}(\mathcal{F}(x))  \right\|^2],
\end{aligned}
\end{equation}
where $p$ stands for the likelihood. As illustrated in Figure \ref{fig:daa}, the decoder, adversarial prediction head, and nuisance prediction head are discarded after the disentangled feature learning process of DAA. The weight of the encoder is then frozen for feature extraction, and a task prediction head with random weight initialization is placed on top of the encoder for specific downstream tasks. Based on their previous work~\cite{han2020disentangled}, Han \textit{et al.} later proposed a soft version of the latent representation disentanglement~\cite{han2021universal}.

\section{Summary of Common Datasets}
\label{sec:dataset}
We summarize common EMG datasets~\cite{pizzolato2017comparison,atzori2016deep,palermo2017repeatability,cote2019deep,amma2015advancing,du2017surface,kanoga2021semi,liu2008muscle,cote2021transferable} that could be used for transfer learning and provide dataset statistics in Table \ref{tab:dataset}, including task category, number of subjects, number of recording device channel, sampling frequency, number of gesture classes, and corresponding citations.

\begin{table*}[!htbp]
% \small
\caption{Summary and statistics of common EMG datasets for transfer learning.}
\label{tab:dataset}
\centering
\resizebox{\textwidth}{!}{%
% \begin{sideways}
\begin{tabular}{ccccccc}
\hline
Dataset &
  Category &
  \# Subject &
  \# Channels &
  Sampling Frequency &
  \# Class &
  Citation \\ \hline
In-house Data &
  - &
  - &
  - &
  - &
  - &
  \tabincell{l}{\cite{li2021improvement}, \cite{lin2020normalisation}, \cite{ameri2019deep}, \cite{jiang2021data},\cite{kanoga2018assessing}, \cite{paassen2018expectation}, \\ \cite{7302056}, \cite{prahm2019counteracting}, \cite{prahm2017transfer}. \cite{6985518}, \cite{gunay2019transfer},\cite{bao2021inter},\\ \cite{tam2021intuitive}, \cite{kobylarz2020thumbs},\cite{zakia2021force}, \cite{sloboda2021utility}, \cite{rezaee2022hybrid},\cite{bird2020cross}, \\\cite{chen2020hand},\cite{soselia2022lower},\cite{zanini2020parkinson},\cite{suri2018transfer},\cite{campbell2021deep} } \\
  NinaPro DB1~\cite{pizzolato2017comparison} &
  Hand Gesture Recognition &
 27 &

  10 &
  100 HZ &
  53 &
  \tabincell{l}{\cite{zhang2021feature}, \cite{ketyko2019domain}, \cite{yu2021surface}} \\
  NinaPro DB2~\cite{pizzolato2017comparison} &
  Hand Gesture Recognition &
 40 &

  12 &
  2000 HZ &
  50 &
  \tabincell{l}{ \cite{zou2021transfer}, \cite{lehmler2021deep}, \cite{kim2019subject}, \cite{zhang2022transductive}, \cite{fan2021improving}, \cite{tsinganos2021transfer}, \\ \cite{marano2021questioning}, \cite{rahimian2021few}, \cite{zanini2020parkinson}, \cite{zhai2017self}} \\
  NinaPro DB3~\cite{pizzolato2017comparison} &
  Hand Gesture Recognition &
  11 &

  12 &
  2000 HZ &
  50 &
  \tabincell{l}{ \cite{zou2021transfer}, \cite{lehmler2021deep}, \cite{kim2019subject},\cite{zhang2022transductive}, \cite{fan2021improving}, \cite{marano2021questioning}, \\\cite{zhai2017self} } \\
  NinaPro DB4~\cite{atzori2016deep} &
  Hand Gesture Recognition &
  10 &

  12 &
  2000 HZ &
  53 &
  \tabincell{l}{ \cite{zou2021transfer}, \cite{lehmler2021deep}} \\
NinaPro DB5~\cite{atzori2016deep} &
  Hand Gesture Recognition &
  10 &

  16 &
  200 HZ &
  53 &
  \tabincell{l}{\cite{kanoga2022subject}, \cite{cote2019deep}, \cite{zou2021transfer}, \cite{zhang2021feature}, \cite{li2021transfer}} \\
NinaPro DB6~\cite{palermo2017repeatability} &
Hand Gesture Recognition &
10 &

16 &
2000 HZ &
8 &
\tabincell{l}{ \cite{zou2021transfer}, \cite{marano2021questioning}} \\

Myo Dataset~\cite{cote2019deep}&
Hand Gesture Recognition &
36 &

8 &
200 HZ &
7 &
\tabincell{l}{\cite{cote2019deep}, \cite{cote2017transfer}} \\

CSL-HDEMG~\cite{amma2015advancing}&
Hand Gesture Recognition &
5 &

192 &
2048 HZ &
27 &
\tabincell{l}{\cite{du2017surface}, \cite{zhang2021feature}} \\
CapgMyo Database~\cite{du2017surface}&
Hand Gesture Recognition &
23 &

128 &
1000 HZ &
7 &
\tabincell{l}{\cite{du2017surface}, \cite{zhang2021feature}, \cite{ketyko2019domain}, \cite{yu2021surface}} \\
SS-STM~\cite{kanoga2021semi}&
Hand Gesture Recognition &
25 &

8 &
200 HZ &
22 &
\tabincell{l}{\cite{hoshino2022comparing}} \\
Multiple Speed Walking~\cite{liu2008muscle}&
Muscle Force Prediction &
8 &

8 &
1080 HZ &
- &
\tabincell{l}{\cite{dao2019deep}} \\
Physical Action Dataset &
Physical Action Recognition &
4 &

8 &
10000 HZ &
20 &
\tabincell{l}{\cite{demir2019surface}} \\
LongTermEMG~\cite{cote2021transferable} &
Hand Gesture Recognition &
20 &

10 &
1000 HZ &
11 &
\tabincell{l}{\cite{cote2021transferable}, \cite{cote2020unsupervised}} \\

\multicolumn{1}{l}{}
\end{tabular}
% \end{sideways}
}
\end{table*}

\section{Discussion and Future Directions}
\label{sec:discussion}
In this section, we revisit EMG transfer learning approaches based on our categorization and discuss the advantages and drawbacks of each category. Given our discussion, we further point out future directions. 
\paragraph{Instance Weighting:} By applying the weight onto the data samples from the source domain, instance weighting makes use of existing source domain data to augment the target domain data to enlarge the size of the data to train the model. This line of method alleviates data shortage when the target domain data are limited. One potential drawback of such methods is that the overall performance is highly dependent on the weighting mechanism and that the target model could suffer from poorly selected and weighted samples from the source domain.
\paragraph{Linear Feature Transformation:} Linear feature transformation based approaches are the most bio-inspired transfer learning approaches of all categories in the sense that the generation of EMG and the recording of EMG could all be abstracted with linear assumption. This line of work is simple and computationally light since the transfer process is simply done by applying a linear transformation on either the data or feature, which is easily done by matrix multiplication. We argue that the linear assumption holds for the transfer scenarios, which are electrodes shift correlated. We mentioned in Section \ref{sec:emg_basics} that certain non-linear factors such as the filtering effect of muscle and fat tissues and muscle fiber recruitment patterns vary across subjects. These non-linear factors could not be modeled with a linear transformation. However, if the underlying subject and recording devices remain the same, electrode shift can then be somewhat captured by such approaches.
\paragraph{Non-linear Feature Transformation:} The non-linearity of this line of work mainly comes from the non-linear activation functions of DNNs. Consequently, the non-linear factors such as subject variation can be modeled in a black-box fashion. Meanwhile, such methods also share the common advantages of DNNs, such as robust feature extraction ability. One main drawback is that DNN based non-linear transformation lacks interpretability in that it's not clear what features are exactly extracted to reduce data distribution discrepancy. Therefore, it's hard to further improve the algorithm since no biological sound clue resides behind the design of the architecture.
\paragraph{Parameter Fine-tuning:} Fine-tuning as transfer learning is simple in practice, since the only operation is to run the training process again on the target domain dataset. However, if the data size of the target domain is limited, the resulted model might suffer from \textit{over-fitting}. Moreover, fine-tuning, in general, suffers from \textit{catastrophic forgetting} where the learned knowledge from the source domain is quickly forgotten with the introduction of new target domain data.
\paragraph{Parameter Sharing:} Parameter sharing based approaches are quite similar to fine-tuning, however, partial network parameters are shared between the source and the target model. By doing so, the aforementioned \textit{catastrophic forgetting} could be alleviated since certain knowledge is considered kept by sharing the associated network parameters. The common practice would be to share the parameters of the feature extractor and to train a task-relevant prediction head from scratch. Freezing the backbone is a common practice when the source domain is believed to be of large size and of similar distribution to the target dataset. Otherwise, there is no guarantee that only training a small fraction of parameters would yield a good transfer performance.
\paragraph{Model Ensemble:} Directly combining data of multiple domains might lead to the neural network not converging smoothly due to data distribution differences. Building individual models with respect to individual domains and then ensembling them best preserves the information for each domain. Since we assume that data distributions from different sessions or subjects vary greatly for EMG applications, thus model ensemble gains the most performance improvement by promoting the diversity of the models. The model ensemble is computational and memory expensive, given that multiple models are stored in memory, and data point is processed multiple times for the final prediction.
\paragraph{Model Structure Calibration:} Existing model structure calibration based models are mainly based on random forest, which in essence is model ensemble already. Thus, this line of work shares the advantages with model ensemble based methods. The structure calibration refers to the growing or pruning operations of individual decision trees. One drawback is that features need to be extracted manually, which is also the drawback of the decision tree itself. It would also be interesting to explore the possibility of calibrating the model structure of DNNs using neural network structure searching tools such as Neural Architecture Search (NAS).
\paragraph{Label Calibration:} This line of work use the source model to label unseen. The labeled and calibrated target domain label is then used to update the model. One advantage is that transferring mechanism of these methods is very in favor of real-world applications. Such methods do not require an expert for target domain data labeling. The transferring process could be deployed on end devices and be automatically applied with new incoming data with a simple user interface. However, since the source domain model label data with knowledge learned from the source domain and will assign label to data points even with previous unseen categories, the label calibration procedure may potentially introduce label noise.
\paragraph{Data Generation:} Generating synthetic EMG data could avoid the tedious workload of data collection and annotation. Given that EMG collection and labeling is very time consuming and requires expertise, generated data of good quality could enhance practicality. However, unlike the data generation in the vision or language community, where the quality of the generated images or texts could easily be verified by human observation, it is hard to evaluate the quality of EMG signals generated. As a consequence, using poorly generated data as data from another domain may bring a negative impact.
\paragraph{Meta/Adversarial Learning Based:} Adversarial learning learns features that are domain irrelevant. Meta learning mimics consecutive transfer learning during the training time so that the model can be adapted to a new domain with limited data. All related methods will perform well on a series of transfer learning with many new target domains. However, the training process of these approaches is either complex or/and introduces additional network components during transferring, which makes it almost impossible for fast transfer learning on an end device.

The essence of EMG transfer learning is to boost the viability of existing machine learning based EMG applications. Consequently, the transfer learning algorithm should bear the following characteristics:
\begin{itemize}
\item[1)] \textbf{Bio-Inspired.} The working mechanism of muscles is relatively well studied and straightforward compared to that of the brain. We point out that the activation patterns of the muscles, relative location between muscles and electrodes, and individual biological characteristics should be explicitly modeled into the neural network to embed the network with A priori knowledge. AlphaFold~\cite{jumper2021highly} is a successful attempt at protein structure prediction with protein A priori knowledge guided network structure design.
\item[2)] \textbf{Hardware-friendly.} Ideally, the re-calibration should be done on end devices rather than on cloud servers. With wearable or even implantable devices, the memory and computation resources are highly restricted. Most current DNN based transfer learning approaches fail to take the hardware constraints into consideration. Future works should incorporate a hardware resource perspective into algorithm design (hardware-software co-design). 
\item[3)] \textbf{User-friendly.} The transferring process should be fast and light in the sense that there should be no heavy data collection procedure that requires user participation. Future works thus should put more attention on transfer learning algorithms that work with limited target domain data and annotation. For instance, given a hand gesture classification task with more than 20 classes, the algorithm is considered user-friendly if the user is required to perform the most simple gesture once for system re-calibration.
\end{itemize}
\section{Acknowledgement}
\label{sec:acknowledgement}
The authors would like to acknowledge start-up funds from Westlake University to the Center of Excellence in Biomedical Research on Advanced Integrated-on-chips Neurotechnologies (CenBRAIN Neurotech) for supporting this project.  The Zhejiang Key R\&D Program Project No. 2021C03002 and the Zhejiang Leading Innovative and Entrepreneur Team Introduction Program No. 2020R01005 both provided funding for this work.

\section*{References}
\bibliographystyle{iopart-num}
\bibliography{ref}

\end{document}